\documentclass[a4paper,11pt]{article}
\usepackage{epsf}
\usepackage{bigints}
\usepackage{stmaryrd}
\usepackage{graphics}
\usepackage{pstricks}
\usepackage{color}
\usepackage{soul}
\usepackage{latexsym}
\usepackage{amsfonts}
\usepackage{amsmath}
\input xy
\xyoption{all}
\makeindex
\date{ }
\def \epsilon {\varepsilon}
\def\norm#1{\left |#1\right |}
\def\Norm#1{\left \|#1\right \|}
\def\upsilon{V}

\title{Integrability of close encounters in 
  the spatial restricted three--body problem
  \footnote{ Preprint version submitted for publication in
    Communications in Contemporary Mathematics,
    DOI: 10.1142/S0219199721500401, $\copyright$ World Scientific
    Publishing Company, https://www.worldscientific.com/worldscinet/ccm}
}

\begin{document}

\author{Franco Cardin and Massimiliano Guzzo
\\ \footnotesize Universit\`a degli Studi di Padova
\\ \footnotesize Dipartimento di Matematica ``Tullio Levi-Civita''
\\ \footnotesize Via Trieste, 63 - 35121 Padova, Italy 
\\ \footnotesize cardin@math.unipd.it, guzzo@math.unipd.it}

\maketitle

\begin{abstract} \par \noindent
We extend to the  spatial case 
a technique of integration of the close encounters
 formulated by Tullio Levi-Civita for the planar restricted three-body
 problem.  We consider the  Hamiltonian introduced in the 
Kustaanheimo-Stiefel regularization and construct a complete integral
of the related Hamil\-ton-Jacobi equation by means of a series convergent in a neighbourhood of the collisions with the primary or secondary body.
\end{abstract}

\section{Introduction}

\noindent
{\bf A.} {\it Motivations.} {Two cornerstone models of physics are the two-body and the three-body problems, which share the same physics --Newton's law of gravitation-- but have a completely different 
mathematical development.}   While in the two-body problem all the orbits are 
classified according to the values of its constants of motion, in the 
three-body problem the constants carry out only a reduction to an 
integral manifold. Poincar\'e worked out the complexity of the dynamics on 
these manifolds; let us mention the deep fundamental results about the non-existence of global analytic first integrals and the non predictability due to homoclinic chaos \cite{Poinc}. An additional issue to the 
 complexity of the three-body problem is due to the gravitational 
singularities, which not only involve collision solutions, 
but also close encounters.  There is a rich recent literature 
about the complex dynamics of the three--body problem whenever close encounters
are concerned, with strong links to the dynamics of comets, near-earth asteroids, and space mission design; see for example \cite{Simo99,FNS,GKZ,Conley67,JM99,Valsecchi,KLMR07,Lega11,GomKoLoMaMasRoss,GL13,celletti1,GL16,GL17,GL18,PG20}. 
In particular, an individual close encounter is sufficient to produce
a resonance transition and 
sequences of such transitions produce orbits which are unpredictable 
on time scales which are very short compared to the secular ones. These
peculiar orbits of the three-body problem are typically observed for 
the comets of the Jupiter family, characterized by resonant transitions
occurring at the close encounters with Jupiter (see paragraph E for
more details).  The numerical integration 
of these orbits is highly critical, due to the strong amplification
of the separation of nearby solutions occurring at each close encounter. 
Therefore, it is essential to approximate analytically the arcs of solutions 
of the three-body problem passing close to a gravitational singularity as much precisely as possible. For the spatial case this problem was formulated by Tisserand (see the Remark, and 
paragraph E). Our paper is about this issue, which has been solved for the 
planar circular restricted three-body problem by Levi-Civita \cite{LC1904,LC1906}.

 More precisely, consider the circular restricted 
three-body problem defined by the motion of a body $P$ of infinitesimally 
small mass 
in the gravitation field of two massive bodies $P_1$
and $P_2$, the primary and secondary body respectively, 
which rotate uniformly around their common center of mass. In a rotating frame,  the Hamiltonian of the problem is:
\begin{equation}
h(x,y,z,p_x,p_y,p_z)= {p_x^2+p_y^2+p_z^2\over 2} + p_x y -p_y x 
-{1-\mu \over r_1}-{\mu\over r_2}  ,
\label{hambarycentric}
\end{equation}
where $r_1=\sqrt{(x+\mu)^2+y^2+z^2}$ and $r_2=\sqrt{(x-1+\mu)^2+y^2+z^2}$ denote 
the distances of $P$ from $P_1,P_2$; notice that as usual the units of mass,
length and time have been chosen so that the masses of $P_1$ and $P_2$ are  $1 - \mu$ and $\mu$ ($\mu \le 1/2$) respectively, their coordinates  are
$(x_1,0,0)=(-\mu,0,0)$, $(x_2,0,0)=(1-\mu,0,0)$ and their revolution period is $2\pi$. Let us consider first the planar motions.  Levi-Civita performed the integration of the close encounters in the planar circular 
restricted three-body problem through the introduction of a transformation which nowadays bears the name 
of Levi-Civita (LC hereafter) regularization. Explicitly: 
\begin{eqnarray}
  x &= &x_j+ u_1^2-u_2^2 \label{regcla}\\
  y & = & 2u_1u_2 \label{regclbb}\\
dt &=& r_j\ ds \label{dt} ,
\end{eqnarray}  
where (\ref{regcla}), (\ref{regclbb}) are 
equivalent to the complex transformation: 
$$
x+iy=x_j+(u_1+iu_2)^2  ,
$$
while (\ref{dt}) is a parametrization of 
the physical time $t$ into the proper time $s$. 
In the last part of the paper \cite{LC1906} Levi-Civita proved the
existence of a local integral of 
the Hamilton-Jacobi equation of the Hamiltonian
representing the planar circular restricted three-body problem
regularized with (\ref{regcla}), (\ref{regclbb}), (\ref{dt}),  
which we call the Levi-Civita Hamiltonian, 
in a neighbourhood of the collision singularity at $P_j$.
The complete integral is constructed as a series analytic at 
$(u_1,u_2)=(0,0)$, whose coefficients can be explicitly computed 
iteratively up to any arbitrary large order. From this series,  
he proved the existence of a second first integral for the problem, independent of the Hamiltonian, defined in a neighbourhood of  
the collision singularity at $P_j$. Therefore, the local integration of planar 
close encounters has been solved by series\footnote{As a matter of fact, Levi-Civita
  constructed the solution of the Hamilton-Jacobi equation
  only for the collision singularity at $P_1$. Nevertheless, 
  Levi-Civita's argument
 is valid also in a neighbourhood of the singularity at the secondary 
 body $P_2$ with some relevant differencies. For example,  
 we notice that while the
 series at $P_1$ is analytic also in $\mu=0$, the series at $P_2$
 is not.}.  The extension of the Levi-Civita regularization to 
the spatial restricted three-body problem  
has been done by Kustaanheimo and Stiefel \cite{KS64,KS65} many decades after Levi-Civita, but the local integrability of the regularized Hamiltonian, which we call the Kustaanheimo-Stiefel Hamiltonian, at collisions has never been 
addressed. Here, our purpose is precisely to extend to the spatial case the point of view followed by Levi-Civita, thus offering a complete integrability of 
the spatial problem in  a neighbourhood of the collision singularities.

\begin{itemize}

\item[ ] {\bf Remark.}  Obviously, the local integrability of close encounters does not mean that the three-body problem is integrable. These local 
integrations are interesting because they are defined in a neighbourhood of a collision set, and allow to 
solve the (open) problem of close encounters, which we formulate
as follows\footnote{\label{TissF}This problem appeared in the literature, with a 
slightly different formulation, already in an earl paper by
Tisserand about the dynamics of comets: {\it ``Le probl\`eme de la d\'etermination
des grandes perturbations d'une com\`ete par Jupiter revient donc
au suivant, qui est tr\`es simple, au moins par son \'enonc\'e: On donne les 
\'el\'ements 
elliptiques ou paraboliques d'une com\`ete, $a_0,e_0,\varpi_0,\ldots $,
au moment o\`u elle p\'en\`etre dans la sph\`ere d'activit\'e 
de Jupiter. Il faut en d\'eduire les \'el\'ements $a_1,e_1,\varpi_1,\ldots$, 
au moment o\`u elle en sort.''} \cite{tisserand}, pag. 243. For more details
about classic and modern astronomical motivations to the problem of 
close encounters we refer to paragraph E of this Introduction).}.
Let $\sigma$ be 
arbitrarily small; for any motion  
$(x(t),y(t),z(t))$ entering the ball $B_{(x_j,0,0)}(\sigma )\subset {\Bbb R}^3$ 
(centered at $(x_j,0,0)$ of radius $\sigma$)   
at time $t_0$ and leaving it at time $t_1$, express $(x(t_1),y(t_1),z(t_1),
p_x(t_1),p_y(t_1),p_z(t_1))$ as an explicit function of  
$(x(t_0),y(t_0),z(t_0),$ $p_x(t_0),p_y(t_0),p_z(t_0))$. We recall that,
while there is a rich literature about the collision manifolds of $N$-body 
problems, the problem of close encounters is of primal 
importance for astronomical applications such as 
the dynamics of comets, of near-Earth asteroids, 
and modern space mission design (see paragraph E of this Introduction 
for a detailed discussion). 
\end{itemize}

\noindent
{\bf B.} {\it Statement of the main result.} The regularizations of the equations of motion in the spatial 
problems are more complicate than those of the planar problem, see for example, \cite{Moser70}.  As for the 
Levi-Civita regularization, the Kustaanheimo-Stiefel regularization (KS hereafter) is defined by the introduction of a transformation
on the space variables and by a time-reparametrization; but the KS space
transformation is more complicate than the LC space transformation,
since it is a map from a space of redundant variables
$u_1,u_2,u_3,u_4$ to a space of Cartesian variables $q_1,q_2,q_3$. 
In fact, for an algebraic reason that we better explain below, the 
generalization of the space transformation (\ref{regcla}), 
(\ref{regclbb}) to the spatial case is related with the extension
of complex numbers to a space of quaternions. 
Precisely, following \cite{KS64,KS65}, we 
introduce the projection map:
\begin{eqnarray}\label{reg1map}
\pi: {\Bbb R}^4 &\longrightarrow& {\Bbb R}^3\cr
(u_1,u_2,u_3,u_4)&\longmapsto& \pi(u_1,u_2,u_3,u_4)=(q_1,q_2,q_3)  ,
\end{eqnarray}
where $(q_1,q_2,q_3,0)=A(u)u$, and:
\begin{equation}\label{matA}
A(u) = 
\left ( \begin{array}{cccc}
u_1 & -u_2 & -u_3 & u_4 \\
u_2 & u_1 &  -u_4 & -u_3 \\
u_3 & u_4 & u_1 & u_2 \\
u_4 & -u_3 & u_2  & -u_1
 \end{array} \right ) 
\end{equation} 
is a matrix which plays a central role in the KS regularization, it is a  
linear homogeneous function of $u_1,\ldots, u_4$ 
and satisfies $A(u)A^T(u)=\norm{u}^2 {\cal I}$. Matrices with such 
properties exist only for $n=1,2,4,8$ (see \cite{hurwitz}; the relation to quaternions is well exploited in \cite{KS64,KS65,cell00,waldvogel,elbialy}). For example, 
for $n=2$ the matrix:
\begin{equation*}
A_2(u) = 
\left ( \begin{array}{cc}
u_1 & -u_2 \\
u_2 & u_1 \\
 \end{array} \right ) 
\end{equation*}
defines the transformation (\ref{regcla}),(\ref{regclbb}) through
$(x-x_j,y)=A_2(u)u$, and the lack of such a matrix for $n=3$ is the reason 
for the definition 
of the KS regularization in a 4-dimensional space.  
Then, for any 
motion in the KS variables we introduce 
the parametrization of time (\ref{dt}); notice that, again, 
we have $r_j=\norm{u}^2$. 
The space and time transformations (\ref{dt}), (\ref{reg1map})
have been used to represent the regularized equations of 
motions of the spatial circular restricted three-body problem
in various forms (see \cite{cell00} for a review of the subject,
and Section 2 for a revisitation).

To better accomplish the technique of integration introduced in \cite{LC1906} 
we first perform the phase-space translation
\begin{equation}
X=x-x_j\ \ ,\ \ Y=y\ \ ,\ \ Z=z\ \ ,\ \ P_X=p_x\ \ ,\ \ 
P_Y=p_y-x_j\ \ ,\ \ P_Z=p_z  ,
\label{planetoXYZ}
\end{equation}
conjugating $h$ to the Hamiltonian (to fix ideas we present all
these computations for $j=2$, so that the reference system defined 
above will be called planetocentric):
$$
H(X,Y,Z,P_X,P_Y,P_Z)= {P_X^2+P_Y^2+P_Z^2\over 2} + P_X Y -P_Y X-{\mu\over
\sqrt{X^2+Y^2+Z^2}}
$$
\begin{equation}
-(1-\mu)\left ({1 \over \sqrt{(X+1)^2+Y^2+Z^2}}-1+X\right )
-(1-\mu)-{(1-\mu)^2\over 2}  ,
\label{hamplanetocentric}
\end{equation}
the constant terms being kept for comparison with the values of the 
original Hamiltonian $h$. The KS regularization is obtained 
from the space transformation (\ref{reg1map}) with $(q_1,q_2,q_3)=(X,Y,Z)$
and, in Section 2, we show that it can be formulated in the following 
Hamiltonian form:
$$
{\cal K}(u,U;E)=
{1\over 8}\norm{U-{ b}_{(0,0,1)}(u)}^2 
-{1\over 2}\norm{u}^2\norm{(0,0,1) \times \pi(u)}^2-\norm{u}^2E_\mu-\mu 
$$
\begin{equation}
-(1-\mu)\norm{u}^2\left ({1 \over \norm{\pi(u)+(1,0,0)}}-1
  +\pi(u)\cdot (1,0,0)\right ) ,\ \ \label{KSham}
\end{equation}
where $U=(U_1,U_2,U_3,U_4)$ denote the conjugate momenta 
to $u=(u_1,u_2,u_3,u_4)$, the vector potential ${ b}_{ \omega}(u)$  (in (\ref{KSham}) 
we have $ \omega=(0,0,1)$), is defined by 
\begin{equation}
{ b}_{ \omega}(u)= 2 A^T(u)\Lambda_{ \omega}A(u)u ,
\ \ \ \ 
\Lambda_{ \omega} = 
\left ( \begin{array}{cccc}
0 & -\omega_3 & \omega_2 & 0 \\
\omega_3 & 0 &  -\omega_1 & 0 \\
-\omega_2 & \omega_1 & 0 & 0 \\
0 & 0 & 0  & 0
 \end{array} \right ) ,\label{vectorpotential}
\end{equation}
and:
$$
E_\mu=E+(1-\mu)+{(1-\mu)^2\over 2} .
$$
The Hamiltonian ${\cal K}(u,U;E)$ is a regularization of the spatial three-body
problem at $P_2$. This means that the solutions $(u(s),U(s))$ of 
the Hamilton equations of ${\cal K}(U,u;E)$ with initial conditions
satisfying:
\begin{itemize}
\item[(i) ] $u(0)\ne 0$;

\item[(ii) ]$l(u(0),U(0))=0$, where  
  \begin{equation}
    l(u,U)=u_4 U_1-u_3 U_2 +u_2 U_3-u_1 U_4\label{bilinearlagran}
    \end{equation}
is called the {\it bilinear form};

\item[(iii) ] ${\cal K}(u(0),U(0);E)=0$,
\end{itemize}
are conjugate, for $s$ in a small neighbourhood of $s=0$, via equations 
(\ref{dt}), (\ref{reg1map}) to solutions
$(X(t),Y(t),Z(t),P_X(t),P_Y(t),P_Z(t))$ of the Hamilton equations of (\ref{hamplanetocentric}). 

Our integration of the close encounters in the spatial circular restricted 
three-body problem
is established on the construction of a complete integral $W(u,\nu;E,\mu)$ 
of the Hamilton-Jacobi equation of ${\cal K}(u,U;E)$, 
defined for all the values of the parameters
$\nu=(\nu_1,\ldots ,\nu_4)$ in a neighbourhood of the
sphere $\norm{\nu}=1$, and analytic in a neighbourhood of
$u=0$. Our main result is the following:\footnote{Theorem 1,  
and Theorem 2 below, have been announced in \cite{cgann}.}
\vskip 0.2 cm
\noindent
    {\bf Theorem 1.} {\it For fixed values of $E_*$ and of $\mu_*>0$, 
      there exists a complete integral $W(u,\nu;E,\mu)$ of the
      Hamilton-Jacobi equation:
\begin{equation}
{\cal K}\left (u,{\partial W\over \partial u}(u,\nu;E,\mu);E\right )= 
\mu (\norm{\nu}^2-1)
\end{equation}
depending on the four parameters $\nu$ and on $E,\mu$, which is analytic for $E,\mu,\nu$ in the set:
$$
\{\norm{\mu-\mu_*}< a ,\ \ \norm{E-E_*}<b ,\ \ \norm{\norm{\nu}-1}< c\}
$$
and $u$ in the (complex) ball:
$$
\{ u\in {\Bbb C}^4: \norm{u}<d\}
$$
with suitable constants $a,b,c,d>0$ (depending only on $E_*,\mu_*$). 
The coefficients of the Taylor
expansions of $W$ with respect to the variables $u$ can be explicitly computed iteratively to any arbitrary order; in particular we have:
\begin{equation}
W= \sqrt{8\mu}\sum_{j=1}^4\nu_j u_j +{\cal O}_3(u)  .
\label{Wfirst}
\end{equation}
}
\noindent
The complete integral $W$ of the Hamilton-Jacobi equation will be used 
to define a canonical transformation:
$$
(n,\nu)=(\hat n(u,U;E,\mu),\hat \nu (u,U;E,\mu))
$$
through the system 
\begin{eqnarray}
U_\ell &=& {\partial W\over \partial u_\ell}(u,\nu;E,\mu),\ \ \ell=1,\ldots ,4 \label{eqU}\\
n_\ell &=& {\partial W\over \partial \nu_\ell}(u,\nu;E,\mu),
\ \ \ell=1,\ldots ,4. \label{eqbeta}  
\end{eqnarray}
conjugating ${\cal K}(u,U;E)$ to the Hamiltonian:
$$
\hat {\cal K}(n,\nu)=\mu (\norm{\nu}^2-1)  .
$$
Therefore, the solutions $(u(s),U(s))$ of the Hamilton equations of
${\cal K}(u,U;E)$ are obtained from the equation:
\begin{equation}
(n(0)+2\mu\ \nu(0)s,\nu(0))=(\hat n(u(s),U(s);E,\mu),\hat \nu (u(s),U(s);E,\mu ) )  .
\label{solutions}
\end{equation}
Formula (\ref{solutions}) provides all the solutions of the spatial
circular restricted three--body problem in a neighbourhood of the
collision set ${\cal C}_2$.  
\vskip 0.2 cm
\noindent 
{\bf C.} {\it On the proof of Theorem 1.}
The proof of Theorem 1 will be achieved through several steps: first, 
a geometric analysis of the KS Hamiltonian is needed to identify the parameters
$\nu_1,\ldots ,\nu_4$, providing the conserved momenta of Hamiltonian
$\hat K(n,\nu)$; second, an analytic part based on the Cauchy-Kowaleski
theorem is used to provide analytic solutions to the Hamilton-Jacobi equation.   The geometric analysis is the original heart of our proof and is 
completely new with respect to the work of Levi-Civita. In fact, while the 
geometric part required by the planar case is rather simpler, 
for the spatial case we need to represent in the space of the 
variables $(u_1,\ldots ,u_4)$ the rotations of the
euclidean space $(q_1,q_2,q_3)$ with matrices which are in SO(4) and leave invariant the bilinear form. Moreover, the subgroup of SO(4) that we obtain this way must be parameterized by parameters $\nu_1,\ldots ,\nu_4$, constrained to the unit sphere, such that the inversion of the system of equations 
(\ref{eqU}), (\ref{eqbeta}) has no singularities (which
arise if, for example, we parameterize the subgroup with three Euler angles).
The analytic 
part is instead the argument that we extend from the 
planar problem, with an additional care for the global definition 
of the family of particular solutions found. 
\vskip 0.2 cm
\noindent
{\bf D.} {\it Complete integrability in the Cartesian phase-space.} 
An additional interesting question concerns the existence of 
Cartesian local first integrals $F(x,y,z,p_x,p_y,p_z)$ independent of 
$h(x,y,z,p_x,p_y,p_z)$ defined in a 
set ${\cal B}\backslash {\cal C}_j$, where  ${\cal C}_j$ is a collision
set:
$$
{\cal C}_j=\{ (x,y,z,p_x,p_y,p_z): (x,y,z)=(x_j,0,0)\} ,\ \  j=1,2,
$$
and ${\cal B}$ is 
a neighbourhood of ${\cal C}_j$\footnote{Through this paper, 
whenever we will refer to a subset of the 
Cartesian phase-space which is a neighbourhood of the collision set, 
we will precisely refer to a set ${\cal B}\backslash {\cal C}_j$, where 
${\cal B}$ is a neighbourhood of the collision set.}  First, we 
remark that
the existence
of Cartesian first integrals is not granted a priori from the existence
of first integrals of the KS Hamiltonian; for example $\norm{\nu}^2$ 
and $l(n,\nu)$ do not provide, with evidence, Cartesian first integrals.
But neither the momenta $\nu_\ell$ provide Cartesian first integrals. 
The deep reason is that the map $\pi$ has not a global smooth inversion 
defined in a neighbourhood of $q=(x-x_j,y,z)=0$ (see \cite{GL18}, where a similar
problem is addressed for the global definition of chaos indicators for
the spatial three body problem), so it can happen that functions $F(n,\nu)$ which are
first integrals for ${\hat {\cal K}}$ do not define 
global Cartesian smooth functions in any neighbourhood of the collision set ${\cal C}_j$.
Precisely, while we are not able to define Cartesian representatives of $\nu_\ell$, $\ell=1,\ldots ,4$,
which are smooth in a neighbourhood of ${\cal C}_j$, we find that the functions:
\begin{eqnarray}
    N_X &=&\nu_1 n_4-\nu_4 n_1\cr 
    N_Y &=&{1\over 2}(\nu_1 n_3-n_1 \nu_3+n_2\nu_4-n_4\nu_2)\cr 
    N_Z &=&{1\over 2}(\nu_1 n_2-n_1 \nu_2+n_4\nu_3-n_3\nu_4)
\end{eqnarray}
    are first integrals and have Cartesian representatives 
${\cal N}_X,{\cal N}_Y,{\cal N}_Z$ globally defined and smooth 
in a neighbourhood of the collision sets. We consider the set of three first integrals:
$$
\Big (\ H \ \ , \ \ {\cal N}^2:={\cal N}_X^2+{\cal N}_Y^2+{\cal N}_Z^2\ \ ,\ \ 
{\cal N}_Z\ \Big )
$$
We notice that, since 
${\cal N}^2,{\cal N}_Z$ are first integrals, we have:
$$
\{ H, {\cal N}^2\} =0 \ \ ,\ \ \{ H,{\cal N}_Z\} =0  .
$$
The Poisson bracket $\{ H,{\cal N}_Z\} =0$ 
is sufficient to grant the 
complete integrability of the planar circular restricted three-body problem 
in a neighbourhood of its collision singularities. It remains 
to understand if even the spatial problem is completely integrable. 
At this regard, we notice that in the space of the variables
$n,\nu$, we have:
\begin{equation}
\{N^2,N_Z\} = l(n,\nu)a(n,\nu)\ \ ,\ \ N^2=N_X^2+N_Y^2+N_Z^2  ,
\label{atipicppint}
\end{equation}
so that the two integrals are in involution on the level set $l(n,\nu)=0$. 
The atypical Poisson bracket in (\ref{atipicppint}) seems a 
rule for the KS regularization. For example, the elementary Poisson 
brackets of $q=\hat q(u),p=\hat p(u,U)$ defined from $\hat q(u)=\pi(u)$,
$(\hat p_1,\hat p_2,\hat p_3,0)={1\over 2\norm{u}^2}A(u)U$, satisfy:
\begin{equation}\label{ppqpint}
\{ \hat q_i, \hat  p_j\} = \delta_{ij} ,\ \ \{\hat  q_i,\hat  q_j\}=0 ,\ \ \{\hat p_i,\hat p_j\}=
l(u,U)\phi_{ij}(u,U),\ \ i,j=1,2,3 .
\end{equation}
From (\ref{atipicppint}) and (\ref{ppqpint}) we will prove the following:
\vskip 0.4 cm
\noindent
{\bf Theorem 2.} {\it The set of first integrals $(H,{\cal N}^2,{\cal N}_Z)$ 
is complete.} 
\vskip 0.4 cm
\noindent
{\bf E.} {\it Astronomical motivations: close encounters.} Astronomers 
were faced with the problem of close encounters few years after the publication 
of Newton's {\it Philosophiae Naturalis Principia Mathematica}, to 
understand the motion of comets.  
Comets are visible from Earth when they are close to the Sun,
therefore apparitions at different epochs correspond to the same comet if they
are linked by the same orbit. While Newton's theory allowed Halley and Clairaut 
to link the former apparitions  of 1531, 1607, 1682 of Halley's comet  
and predict its return for 1759, the dramatic effect 
of close encounters became more evident with the discovery in 1770
of the Lexell's comet. 
Despite the orbit of Lexell's comet was elliptic with period of about 
5.6 years, the comet was not seen in the next 10 years (not either afterwards). Lexell recognized that the comet had likely never had been seen before,
because of a close encounter with Jupiter in 1767, and maybe
it would be never be seen again because of another close encounter 
estimated for 1779. By studying the possible orbits of the comet 
after the latter close encounter, Le Verrier found that the future orbit of the comet was unpredictable \cite{Leverrier}. The method used by Le Verrier
was very modern, since he tried to reproduce the 
orbit of the comet by linking the orbits of the 
two different Sun-comet and Jupiter-comet two-body problems; for
 deep enough close encounters with Jupiter, the linkage 
 expands the small experimental errors in the measure 
of the orbital parameters to complete indetermination. 
Tisserand, who was among the first ones to remark the need of a 
mathematical explanation of the problem (see footnote \ref{TissF}), 
found an approximate integral of motion constraining 
the possible large variations of the orbital parameters \cite{tisserand}
as the effect of a close encounter, but still the problem remained highly undetermined. More recently, \"Opik (\cite{opik}, see \cite{Valsecchi} for a 
recent revisitation) developed
Le Verrier's method and formulated 
a more refined predictive theory of close encounters which, despite the 
good agreement with numerical integrations, still needs a  
mathematical justification (see also \cite{tommei}). 
The short-term indeterminism in the orbit of 
Lexell comet is not an exception, but is typical of comets having fast close 
encounters with the planets. For example, a deep close encounter with 
Jupiter which occurred in 1959 is 
responsible of the indetermination of the past orbit of comet\footnote{Comet
67P/Churyumov-Gerasimenko has been the 
target of the recent mission Rosetta.} 67P which can be 
obtained from backward numerical integrations, for epochs exceeding few 
centuries \cite{GL16,GL17}. Modern topics of cometary dynamics 
where close encounters are relevant raised in the investigations about 
 the formation of solar system. In the modern picture of the Solar system
 there is a population of icy bodies outside Neptune's orbit, of relatively 
small eccentricities and inclinations, which is potentially a reservoir
of periodic comets (see, for example, \cite{gladman} and references therein). This picture poses the mathematical problem of
proving that the orbital resonances and chains of close 
encounters with the giant planets reduce
the perihelion distance of these icy objects from values larger than 
Neptune's aphelion to distances shorter than 3 AU, where the body shows 
its cometary activity. Close encounters are important also for astrodynamics, since they are used in the technique of gravity assist 
to change the energy of a spacecraft: interplanetary missions to the 
giant planets have been possible only thanks to close encounters with the 
planets. The results that we prove
in this paper could be exploited in these problems. In fact, by considering
applications of close encounters, we notice that we have two relevant 
spheres: a sphere $B$ where the gravitational interaction
with the body $P_2$ is dominant (for example, this can be identified
using the Hill's sphere),   
and a smaller sphere $B_2(\sigma)\subset B$ where the close 
encounters are integrated by series. 
So, we have  
a spherical neighbourhood of the Planets in the physical space, with 
radius depending on the energy of the incoming orbit, where one can compute the close encounter, or the incoming and outgoing orbits, with any 
needed precision. The crossing of the interspace 
$B\backslash B_2(\sigma)$ between the two spheres and of the 
region complementary to $B$ where the gravitational interaction
with $P_1$ is dominant needs to be studied with perturbation methods, 
such as those used in \cite{FNS,GKZ}.  
\vskip 0.4 cm
The paper is organized as follows. In Section 2 we revisit the
definition of the KS transformation with respect 
to any spatial frame centered at $P_j$ and arbitrarily rotated; Section
3 is devoted to the identification of suitable parameters for
the definition of a complete integral of the Hamilton-Jacobi equation of
${\cal K}(u,U:E)$; in Section 4 we prove the existence of particular solutions
of the Hamilton-Jacobi equation; in Section 5 we prove the existence of
a complete integral, thus proving Theorem 1, and we use it to define a canonical transformation;
in Section 6 we discuss the existence of Cartesian first integrals and we prove Theorem 2;
in Appendix 1 we revisit the integration of close encounters in the planar three-body problem
done by Levi-Civita in \cite{LC1906}; in Appendix 2 we review a basic formulation
of the Cauchy-Kowaleski theorem.

\section{The KS Hamiltonian revisited}

In order to solve the problem of close encounters in the spatial case
we need to introduce the KS transformation with respect 
to any spatial frame centered at $P_j$ and arbitrarily rotated, while 
in the usual KS transformation the Cartesian coordinates
are referred to a rotating spatial frame with $x$ axis containing 
the primaries $P_1,P_2$ and the $z$ axis orthogonal to their orbit plane. 
In addition, we consider also an arbitrary scaling of the coordinates by a 
factor $\lambda>0$; the scaling will be needed to define the parameters 
of the solutions of the Hamilton-Jacobi equation. 
\vskip 0,4 cm
\noindent
    {\bf The Lagrangian formulation in the Cartesian variables.} We start from the Lagrange function of the
    spatial circular restricted three-body problem:
$$
L_C(x,y,z,\dot x,\dot y,\dot z)={1\over 2}(\dot x^2+\dot y^2+\dot z^2)
+\dot y x -\dot x y+{1\over 2}(x^2+y^2)
$$
$$+{1-\mu \over \sqrt{(x+\mu)^2+y^2+z^2}}
+{\mu \over \sqrt{(x-1+\mu)^2+y^2+z^2}}
$$
$$
={1\over 2}(\dot x^2+\dot y^2+\dot z^2)+(\dot x,\dot y,\dot z)\wedge
(0,0,1)\cdot (x,y,z)+{1\over 2}\norm{(0,0,1)\wedge (x,y,z)}^2
$$
\begin{equation}
+{1-\mu \over \sqrt{(x+\mu)^2+y^2+z^2}}
+{\mu \over \sqrt{(x-1+\mu)^2+y^2+z^2}}
\label{lagrangianc}
\end{equation}
and, for any arbitrary matrix  ${\cal R}\in SO(3)$ and any $\lambda>0$, 
we define the coordinates transformation:
\begin{equation}
(x-x_j,y,z)=\lambda{\cal R}{ q} ,
\label{xc}
\end{equation}
where ${ q}= (q_1,q_2,q_3)$ and (to fix ideas)
$x_j=x_2=1-\mu$, which extends to the transformation on the 
generalized velocities: 
\begin{equation}
(\dot x,\dot y,\dot z)=\lambda{\cal R}
\dot { q} .
\label{xdotc}
\end{equation}
By transforming the Lagrangian $L_C$ with (\ref{xc}), (\ref{xdotc}), and by dropping the constants as well as the
terms which are independent on the $q_i$  and linear in the $\dot q_i$ (which do not contribute to the
Lagrange equations) we obtain the Lagrangian:
$$
L(q,\dot q)= {1\over 2}\lambda^2 \norm{\dot q}^2 +
\lambda^2  (\dot q \wedge \omega)\cdot q +{1\over 2}\lambda^2 \norm{\omega \wedge q}^2
+{\mu\over \lambda\norm{q}}
$$
\begin{equation}
+(1-\mu)\left ({1 \over \norm{\lambda q+e}}+\lambda  q\cdot e\right )  ,
\label{lagran}
\end{equation}
where $\omega={\cal R}^T (0,0,1)$, $e={\cal R}^T (1,0,0)$.
\vskip 0,4 cm
\noindent
    {\bf The redundant variables $\bf u_1,\ldots ,u_4$.} Redundant variables are easily introduced
    in the Lagrangian
formalism (see, for example, \cite{AKN2006}). As a first step, 
we compute the function: 
$$
\hat L (u,\dot u)=   L\left (\pi(u), 
{\partial \pi\over \partial u}(u)\dot u\right )
$$
using the formulas:
$$
(q_1,q_1,q_3,0)=A(u)u
$$
$$
(\dot q_1,\dot q_2,\dot q_3,0)
=A(\dot u)u+A(u)\dot u = 2 A(u)\dot u -2(0,0,0,l(u,\dot u)) ,
$$ 
where $l(u,\dot u)$ is the bilinear form defined in (\ref{bilinearlagran}). We obtain:
$$
\hat L(u,\dot u)=2\lambda^2 \norm{u}^2\norm{\dot u}^2 -2 \lambda^2 l(u,\dot u)^2+\lambda^2{b}_{ \omega}(u)\cdot {\dot { u}}
$$
\begin{equation}
+{1\over 2}\lambda^2\norm{\omega \wedge \pi(u)}^2
+{\mu\over \lambda\norm{u}^2}
+(1-\mu)\left ({1 \over \norm{\lambda\pi(u)+e}}+\lambda\pi(u)\cdot e\right )  ,
\label{lagranu}
\end{equation}
where ${ b}_{\omega}(u)$, is the vector potential already defined
in (\ref{vectorpotential}).
  
Let us compare the solutions of the Lagrange equations 
of $\hat L(u,\dot u)$, which we write in the form:
$$
[\hat L]_i(u,\dot u,\ddot u)=0 \ \ ,\ \ \forall i=1,\ldots ,4
$$
where:
$$
[\hat L]_i(u,\dot u,\ddot u)={d\over dt}{\partial L\over \partial \dot u_i}-
{\partial L\over \partial u_i} ,
$$
with the solutions  of the Lagrange equations 
of $L(q,\dot q)$, which we write in the form:
$$
[L]_j(q,\dot q,\ddot q)=0 \ \ ,\ \ \forall j=1,2,3
$$
where:
$$
[L]_j(q,\dot q,\ddot q)={d\over dt}{\partial L\over \partial \dot q_j}-
{\partial L\over \partial q_j} . 
$$
\vskip 0.4 cm
\noindent
{\bf Proposition 1.} {\it If $u(t)$ is a solution 
of the Lagrange equations of $\hat L(u,\dot u)$ with $u(0)\ne 0$, then $q(t)=\pi(u(t))$
is a solution of the Lagrange equations of $L$ as soon as $u(t)\ne 0$.} 
\vskip 0.4 cm
\noindent
{\bf Proof of  Proposition 1.} For any smooth curve $u(t)$, we have:
$$
\underline {[\hat L]}(u(t),\dot u(t),\ddot u(t))
= \left ( {\partial \pi\over \partial u}\right )^T \underline {[L]}
(\pi (u(t)),{d\over dt}\pi (u(t)),{d^2\over dt^2}\pi (u(t)))
$$
where $\underline {[\hat L]}\in {\Bbb R}^4$, $\underline {[ L]}\in {\Bbb R}^3$ 
are the vectors of components $[\hat L]_i,[L]_j$ respectively. 

Since for $u\ne 0$, the Kernel of the matrix ${\partial \pi\over \partial u}^T$ contains only the vector $(0,0,0)$, any solution $u(t)$ of the 
Lagrange equations of $\hat L$ (i.e. satisfying 
$\underline {[\hat L]}=(0,0,0,0)$)  
projects to a solution $q(t)=\pi(u(t))$ 
of the Lagrange equations of $L$ as soon as $u(t)\ne 0$. \hfill $\square$
\vskip 0.4 cm
The Legendre
transform defined by $\hat L$ is not invertible, since the 
quadratic form $2 \norm{u}^2\norm{\dot u}^2 -2 l(u,\dot u)^2$ in the
generalized velocities $\dot u$ is degenerate; therefore the definition of
the Hamiltonian formalism is more tricky than usual. To remove the
degeneracy we consider the modified Lagrangian:
$$
{\cal L}(u,\dot u)= \hat L(u,\dot u)+2 \lambda^2 l(u,\dot u)^2
=2\lambda^2 \norm{u}^2\norm{\dot u}^2 +\lambda^2{ b}_{ \omega}(u)\cdot {\dot { u}}
$$
\begin{equation}
+{1\over 2}\lambda^2\norm{\omega \wedge \pi(u)}^2
+{\mu\over \lambda\norm{u}^2}
+(1-\mu)\left ({1 \over \norm{\lambda\pi(u)+e}}+\lambda\pi(u)\cdot e\right )  ,
\label{lagranu2}
\end{equation}
whose Legendre transform:
\begin{equation}
U={\partial {\cal L}\over \partial u}= \lambda^2 ( 4 \norm{u}^2 \dot u
+b_{\omega}(u)) ,
\label{legendre}
\end{equation}
where $U=(U_1,U_2,U_3,U_4)$ denote the momenta conjugate
to $u=(u_1,u_2,u_3,u_4)$, 
is non-degenerate for $u\ne 0$. 
\vskip 0.4 cm
\noindent
{\bf Proposition 2.} {\it If $u(t)$ is a solution 
of the Lagrange equations of ${\cal L}(u,\dot u)$ with initial
conditions $u(0),\dot u(0)$ satisfying $u(0)\ne 0$ and $l(u(0),\dot u(0))=0$,
then it is also a solution of the Lagrange equations of ${\hat L}(u,\dot u)$
as soon as $u(t)\ne 0$.}
\vskip 0.4 cm
\noindent
Before proving the Proposition, we remark that the Lagrangian
${\cal L}(u,\dot u)$ is invariant with respect
to the one-parameter family of transformations:
\begin{equation}
{ u} \longmapsto {\cal S}^0_\alpha{ u}
\label{symmetrymap}
\end{equation}
where ${\cal S}^0_\alpha\in SO(4)$ is defined by
\begin{equation}
{\cal S}^0_\alpha = 
\left ( \begin{array}{cccc}
\cos\alpha & 0 & 0 & -\sin\alpha \\
0& \cos\alpha &  \sin\alpha & 0 \\
 0 & -\sin\alpha & \cos\alpha & 0 \\
\sin\alpha & 0 & 0  &\cos\alpha
 \end{array} \right ) ,
\end{equation}
whose orbits define the fibers of the projection $\pi$, i.e. 
$\pi(  {\cal S}^0_\alpha {  u})=\pi({  u})$ for all $\alpha$.
Precisely, for all $u,\dot u,\alpha$, we have:
$$
{\cal L}({\cal S}^0_\alpha u,{\cal S}^0_\alpha\dot u)=
 {\cal L}(u,\dot u)  .
$$
As a consequence, by Noether's theorem, the function:
$$
J(u,\dot u)=\left ({d\over d\alpha}{\cal S}^0_\alpha u\right )_{\vert \alpha=0}\cdot
{\partial {\cal L}\over \partial \dot u}=
(-u_4,u_3,-u_2,u_1)\cdot (4\lambda^2\norm{u}^2\dot u+\lambda^2 b_{\omega}(u))
$$
is a first integral for the Lagrange equations of ${\cal L}$. 
Moreover, since: $(-u_4,u_3,-u_2,u_1)\cdot b_{\omega}(u)$ vanishes
identically, then:
$$
J_o(u,\dot u)=\norm{u}^2 l(u,\dot u)
$$
is a first integral for the Lagrange equations of ${\cal L}$.
\vskip 0.4 cm
\noindent
{\bf Proof of  Proposition 2.} Let us consider a solution $u(t)$ of the 
Lagrange equations of ${\cal L}$ with $u(0)\ne 0$ and 
$l(u(0),\dot u(0))= 0$. Since $J_0(u,\dot u)$ is constant 
along the solution, as soon as $u(t)\ne 0$ we have also
$l(u(t),\dot u(t))=0$, as well as $l(u(t),\ddot u(t))=0$. 

We claim that  $u(t)$ solves
also the Lagrange equations of ${\hat L}$. In fact, we have:
$$
[\hat L]_i=
[{\cal L}]_i-2 \lambda^2\left (
{d\over dt}{\partial \over \partial \dot u_i}l^2(u,\dot u)
-{\partial \over \partial u_i}l^2(u,\dot u) \right )
$$
$$
=[{\cal L}]_i
-4 \lambda^2 \left (
{d\over dt} \left (l(u,\dot u) {\partial \over \partial \dot u_i}l(u,\dot u)
\right )
- l(u,\dot u){\partial \over \partial u_i}l(u,\dot u)\right )  
$$
$$
=[{\cal L}]_i
-4  \lambda^2\left (
l(u,\ddot u) {\partial \over \partial \dot u_i}l(u,\dot u)
+ l(u,\dot u) {d\over dt} {\partial \over \partial \dot u_i}l(u,\dot u)
- l(u,\dot u){\partial \over \partial u_i}l(u,\dot u)\right )  
$$
and when computed along the solution $u(t)$ (so that
$ [{\cal L}]_i=0$, $l(u,\dot u)=0$, $l(u,\ddot u)=0$) we have also:
$$
[\hat L]_i(u(t),\dot u(t),\ddot u(t))=0  .
$$
\hfill $\square$
\vskip 0.4 cm
Finally, we remark that for any initial condition $(q(0),\dot q(0))$
with $q(0)\ne 0$ we have the freedom 
of choosing the initial conditions $(u(0),\dot u(0))$ satisfying:
$$
\pi(u(0))=q(0)\ \ ,\ \
   {\partial \pi\over \partial u}(u(0))\dot u(0) = \dot q(0)\ \ ,\ \  
   l(u(0),\dot u(0))=0 .
   $$
In fact, if $l(u(0),\dot u(0))\ne 0$, 
since the Kernel of ${\partial \pi\over \partial u}(u)$ is generated by $\hat u=(u_4,-u_3,u_2,-u_1)$ we have the freedom 
of adding to $\dot u(0)$ a vector $\xi \hat u$ and 
to select $\xi \in {\Bbb R}$ so that:
$$
l(u(0),\dot u(0)+\xi \hat u)=l(u(0),\dot u(0))+\xi \norm{u(0)}^2=0  .
$$
\vskip 0.4 cm
\noindent
    {\bf The KS Hamiltonian.} The  Legendre
    transform (\ref{legendre}), which is invertible for all $\norm{u}\ne 0$,
    conjugates the Lagrangian system defined by ${\cal L}$ to the Hamiltonian
    system with Hamilton function:
$$
K(u,U)= 
{1\over 8 \lambda^2\norm{u}^2}\norm{U-
\lambda^2{  b}_{  \omega}(u)}^2 
-{1\over 2}\lambda^2\norm{\omega \wedge \pi(u)}^2
$$
\begin{equation}
-{\mu\over \lambda\norm{u}^2}
-(1-\mu)\left ({1 \over \norm{\lambda\pi(u)+e}}+\lambda\pi(u)\cdot e\right )  ,
\label{hamiltu}
\end{equation}
where $U=(U_1,U_2,U_3,U_4)$ are the conjugate momenta to $u=(u_1,u_2,u_3,u_4)$. 
Let us compute the bilinear equality $l(u,\dot u)=0$  
in the Hamiltonian formulation; for all $u\ne 0$ we have:
$$
l(u,\dot u)= {1\over 4\lambda^2 \norm{u}^2}l(u,U-\lambda^2 b_\omega(u))=
{1\over 4\lambda^2 \norm{u}^2}(l(u,U)-\lambda^2 l(u, b_\omega(u)) .
$$
Since $l(u, b_\omega(u))=0$ identically, the  bilinear equality $l(u,\dot u)=0$
is equivalent to the condition $l(u,U)=0$.

The Hamiltonian $K(u,U)$ is still singular at $u=0$; to remove
the singularity we perform the iso-energetic reduction. For any value $E$
we introduce the Hamiltonian:
$$
{\cal K}_{\lambda\cal R}(u,U)=\norm{u}^2\left (K(u,U)-E-{(1-\mu)^2\over 2}\right )
$$
$$
=
{1\over 8\lambda^2}\norm{U-\lambda^2{  b}_{  \omega}(u)}^2 
-{1\over 2}\lambda^2\norm{u}^2\norm{\omega \wedge \pi(u)}^2-\mu\lambda^{-1}
-\norm{u}^2\left (E+(1-\mu)+{(1-\mu)^2\over 2}\right )
$$
\begin{equation}
-(1-\mu)\norm{u}^2\left ({1 \over \norm{\lambda\pi(u)+e}}-1+\lambda\pi(u)\cdot e\right )  ,
\label{hamiltun}
\end{equation}
which we call the KS Hamiltonian. 

The solutions $u(s),U(s)$ of the Hamilton equations:
\begin{eqnarray}
  u'_j &=& {\partial \over \partial U_j}{\cal K}_{\lambda\cal R}\cr
  U'_j &=& -{\partial \over \partial u_j}{\cal K}_{\lambda\cal R}\ \ ,\ \
  j=1,\ldots ,4
\end{eqnarray}
with initial
conditions $u(0)\ne 0$ and ${\cal K}_{\lambda\cal R}(u(0),U(0))=0$ are
conjugate by the time transformation:
$$
t(s)=\int_0^s\norm{u(\sigma)}^2d\sigma
$$
to solutions of the Hamilton equations of $K(u,U)$ as soon as $u(s)\ne 0$.
We also notice that ${\cal K}_{\lambda\cal R}(u,U)$ is invariant with
respect to the one-parameter family of transformations 
$$
(u,U)\longmapsto ({\cal S}^0_\alpha u, {\cal S}^0_\alpha U) ,
$$
i.e. we have:
$$
{\cal K}_{\lambda\cal R}({\cal S}^0_\alpha u, {\cal S}^0_\alpha U)=
{\cal K}_{\lambda\cal R}(u,U) .
$$
As a consequence, $l(u,U)$ is a first integral for this Hamiltonian
system. 
\vskip 0.4 cm
We remark that for $\lambda=1,{\cal R}={\cal I}$ the Hamiltonian:
$$
{\cal K}_{\cal I}(u,U)=
{1\over 8}\norm{U-{  b}_{(0,0,1)}(u)}^2 
-{1\over 2}\norm{u}^2\norm{(0,0,1) \wedge \pi(u)}^2-\mu
-\norm{u}^2\left (E+(1-\mu)+{(1-\mu)^2\over 2}\right )
$$
\begin{equation}
  -(1-\mu)\norm{u}^2\left ({1 \over \norm{\pi(u)+(1,0,0)}}-1
  +\pi(u)\cdot (1,0,0)\right )  
\end{equation}
provides an Hamiltonian formulation of
the traditional KS regularization; see, for example,   
\cite{Froeschle1970,cell00} for alternative derivations.

\section{The Hamilton-Jacobi equation for the KS Hamiltonian: the 
parameters space}

Our aim is to define a complete 
integral of the Hamilton-Jacobi equation:
\begin{equation}
{\cal K}_{\cal I}\left (u ,{\partial W\over \partial u}\right )= \kappa ,
\label{HJ-K4}
\end{equation}
which is analytic in a neighbourhood of $u=0$, 
obtained from a family of solutions of (\ref{HJ-K4}) depending on suitable
four parameters. Therefore, we proceed by defining families of particular
solutions $\tilde W$ of the Hamilton--Jacobi equations:
\begin{equation*}
{\cal K}_{\lambda\cal R}\left (u ,{\partial \tilde W\over \partial u}\right )=
\kappa 
\end{equation*}
where ${\cal R}\in SO(3)$ is an arbitrary rotation matrix of
the euclidean three-dimensional space and $\lambda>0$,   
with $\tilde W$ vanishing identically on an hyperplane defined by the choice of 
${\cal R}$. 
\vskip 0.4 cm
\noindent
{\bf Remark.} This procedure depends on four free 
parameters related to $\lambda>0$ and to the
matrix ${\cal R}\in SO(3)$, which in the end will provide the four parameters
needed to define a complete solution of the Hamilton-Jacobi equation.
The first idea to extend the argument of Levi-Civita would seem 
that of using the group $SO(4)$ to transform the KS Hamiltonian 
${\cal K}_{\cal I}$,  and then to define families of particular
solutions $\tilde W$ of the Hamilton--Jacobi equations:
\begin{equation*}
\tilde {\cal K}\left (u ,{\partial \tilde W\over \partial u}\right )= \kappa 
\end{equation*}
where $\tilde {\cal K}(u,U)={\tilde K}_{\cal I}(Su,SU)$ with $S\in SO(4)$,  
with $\tilde W$ vanishing identically on an hyperplane defined by the choice of 
$S$. The problem is that, for arbitrary matrix $S\in SO(4)$, 
the bilinear form $l(u,U)$ is not invariant, i.e. $l(Su,SU)\ne l(u,U)$ on 
some $u,U$. We therefore follow a different strategy.  
\vskip 0.4 cm
\noindent
We have therefore to find a family of transformations on ${\mathbb R}^4$ such
that:
\begin{itemize}
\item[-]  they project on the linear transformations
 of the three--dimensional euclidean space 
\begin{equation*}
(X,Y,Z) \mapsto \lambda {\cal R}(X,Y,Z)
\end{equation*}
 with $\lambda>0$ and
${\cal R}\in SO(3)$;

\item[-] their canonical extensions to the momenta leave invariant 
the diagram about the conjugation of Hamiltonians represented in figure 
1; 

\item[-] their canonical extensions to the momenta leave invariant 
the bilinear form $l(u,U)$ (up to the multiplication with a 
constant different from zero). 
\end{itemize}
We find that the matrices: 
\begin{equation*}
{\cal S}_\nu = 
\left ( \begin{array}{cccc}
\nu_1 & -\nu_2 & -\nu_3 & -\nu_4 \\
\nu_2 & \nu_1 &  -\nu_4 & \nu_3 \\
\nu_3 & \nu_4 & \nu_1 & -\nu_2  \\
\nu_4 & -\nu_3 & \nu_2  & \nu_1
 \end{array} \right ) ,
\end{equation*}
with $\nu=(\nu_1,\nu_2,\nu_3,\nu_4)\in {\mathbb R}^4\backslash 0$,
satisfy:
\begin{equation}
  {\cal S}_\nu{\cal S}_\nu^T = \norm{\nu}^2 {\cal I}  ,
\label{SST}
\end{equation}
and define linear transformations of ${\Bbb R}^4$ which project
on linear transformation of the three-dimensional space so that,
for any $  u\in{\Bbb R}^4$, we have:
\begin{equation}
\pi({\cal S}_\nu u)= {\cal R}_{\nu}\pi(u)
\label{pisr}
\end{equation}
where:
\begin{equation}
{\cal R}_{\nu}=
\left ( \begin{array}{ccc}
\nu_1^2 -\nu_2^2 -\nu_3^2 +\nu_4^2   & -2(\nu_1\nu_2+\nu_3\nu_4) & 
 -2(\nu_1\nu_3-\nu_2\nu_4)\\
2(\nu_1\nu_2-\nu_3\nu_4) & \nu_1^2 -\nu_2^2 +\nu_3^2 -\nu_4^2
 &-2(\nu_2\nu_3+\nu_1\nu_4) \\
2(\nu_1\nu_3+\nu_2\nu_4)  &-2(\nu_2\nu_3-\nu_1\nu_4)  & 
\nu_1^2 +\nu_2^2 -\nu_3^2 -\nu_4^2
 \end{array} \right )  \label{rnu}
\end{equation}
is a matrix satisfying:
\begin{equation}
  {\cal R}_{\nu}  {\cal R}_{\nu}^T=\norm{\nu}^4{\cal I} ,
\end{equation}
which depends on the $\nu_j$ as in
the Euler-Rodrigues formula.

Moreover, for all $(u,U)\in T^*{\Bbb R}^4$, we have:
$$
l({\cal S}_\nu u,{\cal S}_\nu U)=\norm{\nu}^2l(u,U)\ \ .
$$
We therefore consider the set of matrices:
$$
{\cal S}= \cup_{\nu \in {\Bbb R}^4\backslash 0}S_{\nu}
$$
and the map:
$$
\Pi: {\cal S} \longrightarrow SO(3)
$$
$$
S_\nu \longmapsto \Pi(S_\nu)={1\over \norm{\nu}^2}{\cal R}_{\nu}.
$$
The map $\Pi$ is surjective. We have the following:
\vskip 0.4 cm
\noindent
{\bf Proposition 3.} {\it For any matrix $S_\nu\in {\cal S}$ we have the identity:
\begin{equation}
{\cal K}_{\cal I}(S_\nu u,S^{-T}_\nu U)=\norm{\nu}^2{\cal K}_{\norm{\nu}^2\Pi(S_\nu)}(u,U)  .\label{commdiag}
\end{equation}
}
\vskip 0.4 cm
\noindent
    {\bf Proof of Proposition 3.} Let us denote $u=S_\nu\tilde u,U=S_\nu\tilde U$; 
    we have the following identities:
\begin{itemize}
\item[-] $\norm{u}=\norm{\nu}\norm{\tilde u}$;

\item[-] $\norm{\pi(u)+(1,0,0)}=\norm{\pi(S_\nu \tilde u)+(1,0,0)}=
  \norm{\norm{\nu}^2\Pi(S_\nu)\pi(\tilde u)+(1,0,0)}$

\hskip 2 cm$=\norm{\norm{\nu^2}\pi(\tilde u)+\Pi(S_\nu)^T(1,0,0)}$;

\item[-] $\pi(u)\cdot (1,0,0)= \pi(S_\nu\tilde u)\cdot (1,0,0)=
\norm{\nu}^2\pi(\tilde u)\cdot \Pi(S_\nu)^T(1,0,0)$; 

\item[-] $\norm{(0,0,1) \wedge \pi(u)}=\norm{(0,0,1) \wedge 
\pi(S_\nu\tilde u)}=\norm{\nu}^2\norm{(0,0,1) \wedge 
  \Pi(S_\nu)\pi(\tilde u)}$

\hskip 2 cm$=\norm{\nu}^2\norm{\Pi(S_\nu)^T(0,0,1) \wedge \pi(\tilde u)}$,
\end{itemize}
which are proved from (\ref{SST}) and (\ref{pisr}). Finally, we prove:
\begin{equation}
  \norm{S_\nu^{-T}\tilde U-  b_{(0,0,1)}(S_\nu \tilde u)}^2=
  {1\over \norm{\nu}^2}
\norm{\tilde U-\norm{\nu}^4  b_{\Pi(S_\nu)^T(0,0,1)}(\tilde u)}^2  .
\label{eqprop}
\end{equation}
From direct computation, for any $u\in {\Bbb R}^4$, we obtain:
$$
A(S_\nu u)S_\nu = \hat {\cal R}_\nu A(u)
$$
with:
$$
\hat {\cal R}_\nu =\left ( \begin{array}{cccc}
\nu_1^2 -\nu_2^2 -\nu_3^2 +\nu_4^2   & -2(\nu_1\nu_2+\nu_3\nu_4) & 
 -2(\nu_1\nu_3-\nu_2\nu_4)&0\\
2(\nu_1\nu_2-\nu_3\nu_4) & \nu_1^2 -\nu_2^2 +\nu_3^2 -\nu_4^2
 &-2(\nu_2\nu_3+\nu_1\nu_4)&0 \\
2(\nu_1\nu_3+\nu_2\nu_4)  &-2(\nu_2\nu_3-\nu_1\nu_4)  & 
\nu_1^2 +\nu_2^2 -\nu_3^2 -\nu_4^2&0\\
0 & 0& 0 & \norm{\nu}^2
 \end{array} \right )  .
$$
As a consequence, using  (\ref{SST}) and by recalling the definition
(\ref{vectorpotential}) of
the vector potential $  b_\omega$, we have:
$$
\norm{S^{-T}_\nu\tilde U-  b_{(0,0,1)}(S_\nu \tilde u)}^2=
\norm{{1\over \norm{\nu}^2}S_\nu\tilde U-2A(S_\nu \tilde u)^T\Lambda_{(0,0,1)}
  A(S_\nu \tilde u)S_\nu 
  \tilde u}^2
$$
$$
={1\over \norm{\nu}^2}
\norm{\tilde U-2  S^T_\nu A(S_\nu \tilde u)^T\Lambda_{(0,0,1)} A(S_\nu\tilde u)S_\nu\tilde u}^2
$$
$$
={1\over \norm{\nu}^2}
\norm{\tilde U-2 A(\tilde u)^T{\hat {\cal R}}^T_\nu
  \Lambda_{(0,0,1)} \hat {\cal R}_\nu A(\tilde u)\tilde u}^2=
{1\over \norm{\nu}^2}
\norm{\tilde U-2{\norm{\nu}}^4 A(\tilde u)^T{{\hat {\cal R}}^T_\nu\over \norm{\nu}^2}
  \Lambda_{(0,0,1)} \hat {{\cal R}_\nu \over \norm{\nu}^2}
  A(\tilde u)\tilde u}^2
$$
$$
={1\over \norm{\nu}^2}
\norm{\tilde U-\norm{\nu}^4  b_{\Pi(S_\nu)^T(0,0,1)}(\tilde u)}^2
$$
where the last equality is a consequence of the fact that,
for any $\omega\in {\Bbb R}^3$, the 
matrix:
$$
\tilde \Lambda_{  \omega} = 
\left ( \begin{array}{ccc}
0 & -\omega_3 & \omega_2  \\
\omega_3 & 0 &  -\omega_1  \\
-\omega_2 & \omega_1 & 0 
 \end{array} \right ) .
$$
represents the linear transformations of ${\Bbb R}^3$:
$$
\tilde \Lambda_{  \omega} \underline  x= \omega \wedge \underline x
$$
for all $\underline x\in {\Bbb R}^3$; then, for all 
$\underline x\in {\Bbb R}^3$ we have also:
$$
{\Pi(S_\nu)}^T \tilde\Lambda_{(0,0,1)} \Pi(S_\nu)\underline x = 
{\Pi(S_\nu)}^T ((0,0,1) \wedge (\Pi(S)\underline x))=
({\Pi(S_\nu)}^T(0,0,1)) \wedge \underline x ,
$$
and therefore 
$$
{\Pi(S_\nu)}^T_\nu
\tilde \Lambda_\omega \hat \Pi(S_\nu)=\tilde \Lambda_{\Pi(S_\nu)^T(0,0,1)} .
$$
From all the previous equalities we obtain (\ref{commdiag}). \hfill $\square$

\begin{figure} 
  $$
  \xymatrix{
L_{\cal I}(q, \dot q) \ar[r]^{ q={\cal R}\tilde q} \ar[d]^{KS} & L_{\cal R}(\tilde q,\dot {\tilde q}) 
\ar[d]^{KS}\\
{\cal L}_{\cal I}(u,\dot u),\ l(u,\dot u)=0 
\ar[d]^{Legendre} & {\cal L}_{\cal R}(\tilde u,\dot {\tilde u}),\ l(\tilde u,\dot {\tilde u})=0 \ar[d]^{Legendre}\\
{\cal K}_{\cal I}(u,U),\ l(u,U)=0 \ar@{.}[r]^{u=S\tilde u, U=S\tilde U }  & {\cal K}_{\cal R}(\tilde u,\tilde U),\ l(\tilde u,\tilde U)=0 }
  $$
  \caption{For any $S\in {\cal S}$ and ${\cal R}=\Pi(S)$ the 
diagram is commutative. }
\label{fig2}
\end{figure}

\section{The Hamilton-Jacobi equation for the KS Hamiltonian: 
particular solutions}
 
In this Section we prove the existence of particular
solutions $\tilde W$ of the Hamilton-Jacobi equation:
\begin{equation}
{\cal K}_{\norm{\nu}^2\Pi(S_\nu)}\left (u , {\partial \tilde W\over \partial u}\right )={\kappa \over \norm{\nu}^2},
\label{HJequationPS}
\end{equation}
with the following properties:
\begin{itemize}
\item[-] the solutions $\tilde W(u;E,\mu,\kappa,\nu_1,\ldots ,\nu_4)$
are defined for any value of the parameters $(E,\mu,\kappa,\nu_1,\ldots ,\nu_4)$ in a set ${\cal D}_{\alpha,\ldots ,\delta}(E_*,\mu_*)$ defined by fixed values $E_*$ and $\mu>0$, and by suitably small  
$\alpha,\beta,\gamma,\delta>0$:
$$
\norm{\mu-\mu_*}<\alpha
$$
$$
\norm{E-E_*}<\beta
$$
$$
\norm{\kappa}< \gamma
$$
$$
\{ \nu \in {\Bbb R}^4: \ \ 1-\delta < \norm{\nu}< 1+\delta \}  ,
$$
and for any value of the parameters in this set it
is analytic in the same common domain:
$$
u \in {\Bbb C}^4: \ \norm{u}< \sigma  ,
$$
with $\sigma>0$ (depending only on $E_*,\mu_*,\alpha,\ldots ,\delta$). 
\item[-] they satisfy:
\begin{equation}
\tilde W(0,u_2,u_3,u_4;E,\mu,\kappa,\nu_1,\ldots ,\nu_4)=0
\label{ckconditionn2}
\end{equation}
for all $u_2,u_3,u_4$ in a neighbourhood of $0$.

\item[-] they are analytic also with respect to the parameters. 
\end{itemize}
We remark that the domain above considered is local in the variables
$u$ and in the parameters $E,\mu,\kappa$, but is not local
in the parameters $\nu$ which are naturally defined in a 
neighbourhood of ${\Bbb S}^3$. Therefore, since the proof will be obtained
from the Cauchy-Kowaleski theorem, which grants the existence
of local analytic solutions of PDE, we have to pay some care in
proving the global character of the solutions obtained from
the  Cauchy-Kowaleski theorem with respect to the parameters $\nu_i$. 
\vskip 0,4 cm

In order to apply the Cauchy-Kowaleski theorem, we first rewrite the HJ
equation (\ref{HJequationPS}) as follows:
$$
{\partial \tilde W\over \partial u_1}=\norm{\nu}^4
{ b}_{1, \omega}(u)
\pm\sqrt{8}\norm{\nu}\left (\mu+\kappa + 
{1\over 2}\norm{\nu}^6\norm{u}^2\norm{\omega \wedge \pi(u)}^2
+\norm{u}^2\norm{\nu}^2E_\mu \right .
$$
$$
+(1-\mu)\norm{u}^2\norm{\nu}^2\left ({1 \over \norm{\norm{\nu}^2\pi(u)+e}}-1+\norm{\nu}^2\pi(u)\cdot e\right )
$$
\begin{equation}
\left .
-{1\over 8\norm{\nu}^2}\sum_{j=2}^4 \left ( {\partial \tilde W\over \partial u_j}
-\norm{\nu}^4{ b}_{j, \omega}(u)\right )^2\right )^{1\over 2}
\label{eqCK4}
\end{equation}
where $E_\mu= E+(1-\mu)+{(1-\mu)^2\over 2}$, $ \omega=\Pi(S_\nu)^T(0,0,1)$,
$  e=\Pi(S_\nu)^T(1,0,0)$. We solve the previous equation
by selecting the positive sign in front of the square root (the minus
would provide a different solution), and therefore we consider the function:
$$
F(u_1,\ldots ,u_4,p_1,p_2,p_3;E,\mu,\kappa,\nu_1,\ldots ,\nu_4)=
\norm{\nu}^4{  b}_{1,  \omega}(u)
$$
$$
+\sqrt{8}\norm{\nu}\left (\mu+\kappa + 
{1\over 2}\norm{\nu}^6\norm{u}^2\norm{\omega \wedge \pi(u)}^2+\norm{u}^2\norm{\nu}^2E_\mu \right .
$$
$$
+(1-\mu)\norm{u}^2\norm{\nu}^2\left ({1 \over \norm{\norm{\nu}^2\pi(u)+e}}-1+\norm{\nu}^2\pi(u)\cdot e\right )
$$
\begin{equation}
\left .
-{1\over 8\norm{\nu}^2}\sum_{j=2}^4 \left ( p_{j-1}
-\norm{\nu}^4{  b}_{j,  \omega}(u)\right )^2\right )^{1\over 2}
\label{FCK}
\end{equation}
which depends parametrically on $E,\mu,\kappa,\nu_1,\ldots ,\nu_4$. For any fixed $E_*,\mu_*$ with $\mu_*>0$ there exist $\alpha_0,\ldots ,\delta_0$
and $\sigma_0$ such that $F$ is analytic for all 
$(E,\kappa,\nu_1,\ldots ,\nu_4)\in {\cal D}_{\alpha_0,\ldots ,\delta_0}$
in the set $\norm{u}<\sigma_0$.
\vskip 0.4 cm
We first apply the Cauchy-Kovaleskaia theorem to the first-order PDE:
\begin{equation}
{\partial \tilde W\over \partial u_1} = 
F\left (u_1,\ldots ,u_4,{\partial \tilde W\over \partial u_2},
{\partial \tilde W\over \partial u_3},{\partial \tilde W\over \partial u_4};
E,\mu,\kappa,\nu_1,\ldots ,\nu_4\right )
\label{ckfirstorder}
\end{equation}
where $E,\mu,\kappa,\nu_1,\ldots ,\nu_4$ are fixed in
some set ${\cal D}_{\alpha_1,\ldots ,\delta_1}$, with
the boundary condition (\ref{ckconditionn2}):
$$
\tilde W(0,u_2,u_3,u_4;E,\mu,\kappa,\nu_1,\ldots ,\nu_4)=0
$$
for $u_2,u_3,u_4$ in a neighbourhood of $u=0$. We obtain
(see Section \ref{appendixck}) the existence of a unique solution
$\tilde W(u;E,\mu,\kappa,\nu_1,\ldots ,\nu_4)$ of such PDE problem which is  
analytic in a neighbourhood of $u=0$, and the radius of convergence
of the series:
\begin{equation}
\tilde W= \sum_{i_1,\ldots ,i_4\geq 0}c_{i_1,\ldots i_4}(E,\mu,\kappa,\nu)
u_1^{i_1}\ldots u_4^{i_4}
\label{series1}
\end{equation}
is common for all the values of the parameters in the set
${\cal D}_{\alpha_1,\ldots ,\delta_1}$. The coefficients
$c_{i_1,\ldots i_4}(E,\mu,\kappa,\nu)$ can be computed iteratively
in the order $i_1+\ldots +i_4$, and since they are functions globally
defined in ${\cal D}_{\alpha_1,\ldots ,\delta_1}$, the
series (\ref{series1}) is globally defined in the
${\cal D}_{\alpha_1,\ldots ,\delta_1}$. In particular,
we have:
\begin{equation}
\tilde W = \sqrt{8(\mu+\kappa)\norm{\nu}^2}u_1+ {E_\mu\norm{\nu}^3\over \sqrt{\mu+\kappa}}u_1\sqrt{2}
\left ( {u_1^2\over 3}+ u_2^2+u_3^2+u_4^2\right )+u_1{\cal O}_3(u)  .
\label{tildewo4}
\end{equation}
It remains to establish the regularity of the function $\tilde W$ defined
the series (\ref{series1}) with respect to the parameters $E,\mu,\kappa,\nu$. 
Therefore we apply a second time the Cauchy-Kowaleski theorem to the
first-order PDE (\ref{ckfirstorder}) by considering
the independent variables $(u_1,u_2,u_3,u_4,E,\mu,\kappa,\nu_1,\ldots
,\nu_4)$ in a neighbourhood of $(u_1,u_2,u_3,u_4,E,\mu,\kappa,\nu_1,\ldots
,\nu_4)=(0,0,0,0,E_*,\mu_*,0,\nu_1^*,\ldots ,\nu_4^*)$ with
$\nu^*\in {\Bbb S}^3$, with
the boundary condition:
$$
\tilde W(0,u_2,u_3,u_4;E,\mu,\kappa,\nu_1,\ldots ,\nu_4)=0
$$
for $u_2,u_3,u_4$ in a neighbourhood of $u=0$ and for all
$E,\mu,\kappa,\nu$ in a neighbourhood of $E_*,\mu_*,0,\nu^*$. 
 We obtain
(see Section \ref{appendixck}) the existence of a unique solution
$\tilde W_1(u;E,\mu,\kappa,\nu_1,\ldots ,\nu_4)$ of such PDE problem which is  
 analytic in a neighbourhood of $(u,E,\mu,\kappa,\nu)=(0,E_*,\mu_*,0,\nu_*)$,
 with series expansion:
$$
\tilde W_1= \sum_{i_1,\ldots ,i_{11}\geq 0}d_{i_1,\ldots i_{11}}(E_*,\mu_*,\nu_*)
u_1^{i_1}\ldots u_4^{i_4}(E-E_*)^{i_5}(\mu-\mu_*)^{i_6}
\kappa^{i_7}(\nu_1-\nu^*_1)^{i_8}\cdots (\nu_4-\nu^*_4)^{i_{11}}
$$
converging within a radius $\rho(E_*,\mu_*,\nu_*)$ depending
only on $(E_*,\mu_*,\nu_*)$. But since $\tilde W_1$ is also
a solution of the PDE problem where the $E,\mu,\kappa,\nu$ are
given parameters, and $\tilde W_1$ satisfy the same boundary
condition (\ref{ckconditionn2}), from uniqueness we obtain
$$
\tilde W_1(u;E,\mu,\kappa,\nu_1,\ldots ,\nu_4)=
\tilde W(u;E,\mu,\kappa,\nu_1,\ldots ,\nu_4)  ,
$$
and this proves the analyticity of the global solution $\tilde W$
for any value of the parameters in some ${\cal D}_{\alpha,\ldots ,\beta}$
and for some $\norm{u}\leq \sigma$. 

\section{The Hamilton-Jacobi equation for the KS Hamiltonian: 
a complete integral}

Theorem 1 follows from the following:
\vskip 0,4 cm
\noindent
    {\bf Proposition 4.} {\it For fixed values of $E_*$ and of $\mu_*>0$, 
      there exists a complete integral $W(u,\nu;E,\mu)$ of the
      Hamilton-Jacobi equation (\ref{HJ-K4}) depending on the four parameters
      $\nu$ and two additional parameters $E,\mu$, with 
$$
\kappa=\mu (\norm{\nu}^2-1)   .
$$ 
and analytic for $E,\mu,\nu$ in the set:
$$
\{\norm{\mu-\mu_*}< a ,\ \ \norm{E-E_*}<b ,\ \
\nu\in {\Bbb R}^4: \norm{\norm{\nu}-1}< c\}
$$
and $u$ in the (complex) ball:
$$
B_\sigma = \{ u\in {\Bbb C}^4: \norm{\nu}<d\}
$$
with suitable $a,b,c,d>0$. The coefficients of the Taylor
expansions of $W$ with respect to the variables $u$ can be explicitly computed iteratively; in particular
we have:
\begin{equation}
W= \sqrt{8\mu}\sum_{j=1}^4\nu_j u_j +{\cal O}_3(u)  .
\label{Wfirst}
\end{equation}
}
\vskip 0,4 cm
\noindent
{\bf Proof of Proposition 4.} The complete integral is defined by:
$$
W(u;E,\mu,\nu)=\tilde W(\norm{\nu}^{-2}S_\nu^T u;E,\mu,\kappa_\nu,\nu)  ,
$$
with $\kappa_\nu=\mu (\norm{\nu}^2-1)$, 
where
$\tilde W(\tilde u;E,\mu,\kappa,\nu)$ denotes the
solution of  the Hamilton-Jacobi
equation (\ref{HJequationPS}):
\begin{equation}
  {\cal K}_{\norm{\nu}^2\Pi(S_\nu)}\left (\tilde u ,
  {\partial \tilde W\over \partial \tilde u}(\tilde u,E,\mu,\kappa,\nu)
  \right )=
  {\kappa \over \norm{\nu}^2},
\end{equation}
as it has been defined in the previous section. In fact, since we have:
$$
{\partial W\over \partial u}(u;E,\mu,\nu)=
\norm{\nu}^{-2} S_\nu {\partial \tilde W\over \partial \tilde u}(\norm{\nu}^{-2} S^T_\nu u,E,\mu,\kappa_\nu,\nu) ,
$$
using Proposition 1, and setting $u=S_\nu \tilde u$, we obtain
$$
{\cal K}_{\cal I}\left (u,{\partial W\over \partial u}(u;E,\mu,\nu)\right ) =
{\cal K}_{\cal I}\left ( S_\nu {\tilde u}, S^{-T}_\nu
{\partial \tilde W\over \partial \tilde u}(\tilde u,E,\mu,\kappa_\nu,\nu)\right )
$$
$$
= \norm{\nu}^{2}{\cal K}_{\norm{\nu}^2\Pi(S_\nu)}\left ( {\tilde u},
{\partial \tilde W\over \partial \tilde u}(\tilde u,E,\mu,\kappa_\nu,\nu )\right )=\kappa_\nu=\mu(\norm{\nu}^2-1) .
$$
By replacing in (\ref{tildewo4}) $\kappa$ with $\kappa_\nu$ and
$u$ with $\norm{\nu}^{-2} S^T_\nu u$ 
we obtain (\ref{Wfirst}). Therefore, the determinant:
$$
j_4(u,\nu;E,\mu)= \det \left ( 
{\partial  W\over \partial u_i\partial \nu_j}  \right ) 
$$
satisfies:
\begin{equation}
j_4(0, \nu;E,\mu) = 64 \mu^2  .
\label{j4}
\end{equation}
Therefore, $W$ is a complete integral of the Hamilton-Jacobi equation
in a neighbourhood of $u=0$.\hfill $\square$
\vskip 0.4 cm
\noindent
Let us analyze some consequences of Theorem 1.

For any $\nu \in {\Bbb S}^3$, which corresponds to $\kappa=0$,
the function $W$ defines the foliation:
$$
\Gamma_{\nu}=\left \{ (u,U)\in T^*{\mathbb R}^4: \norm{u}<\sigma \ \ ,\ \ 
U_j={\partial W\over \partial u_j}(u;E,\mu,\nu)\right \} ,
$$ 
which is locally invariant (the solutions with initial 
conditions in a leaf $\Gamma_\nu$ can flow out of it in the future and/or 
in the past). Since we are interested in motions of the KS Hamiltonian 
${\cal K}_{\cal I}$ which project on motions of the three--body problem,  
and since the leaves $\Gamma_\nu$ are foliated by the first
integral $l(u,U)$, we consider:
$$
\tilde \Gamma_{\nu}=\left \{ (u,U)\in T^*{\mathbb R}^4: \norm{u}<\sigma \ \ ,\ \ 
U_j={\partial W\over \partial u_j}(u;E,\mu,\nu)\ \ ,\ \ l(u,U)=0\right \} .
$$
{\bf Proposition 5.} {\it For any $\nu\in {\mathbb S}^3$, 
  $\tilde \Gamma_{\nu}$  is a manifold of dimension 3 in a neighbourhood
  of $(u,U)=(0,\sqrt{8\mu}\ \nu)$.}
\vskip 0.4 cm
\noindent 
{\bf Proof of Proposition 5.} The set $\tilde \Gamma_{\nu}$ is obtained from
the solutions $(u,U)$ of the system:
$$
F_1(u,U)=0\ \ ,\ \ F_5(u,U)=0 ,
$$
where:
$$
F_j(u,U)=U_j-{\partial W\over \partial u_j}(u;E,\mu,\nu )\ \ ,j=1,\ldots 4
$$
$$
F_5(u,U)=l(u,U) ,
$$
with $\norm{u}<\sigma$. Since from (\ref{Wfirst}) we have:
$$
F_j=U_j-\sqrt{8 \mu}\nu_j+{\cal O}_2(u)\ \ ,\ \ j=1,\ldots ,4 ,
$$
the restriction of the Jacobian matrix of the map
$F=(F_1,\ldots, F_5)$ to $\tilde \Gamma_\nu$ 
has the representation:
$$
{\cal J}(u,U)_{\vert\tilde \Gamma_\nu} = \left (\begin{array}{ccccc}
{\nabla_u F_1} & {\nabla_u F_2}& {\nabla_u F_3}& {\nabla_u F_4}& {\nabla_u F_5}\\
{\nabla_U F_1} & {\nabla_U F_2}& {\nabla_U F_3}& {\nabla_U F_4}& {\nabla_U F_5}
 \end{array} \right )_{\vert\tilde \Gamma_\nu}
$$
$$
=
 \left (\begin{array}{ccccc}
 0 & 0 & 0 & 0& U_4 \\
 0 & 0 & 0 & 0& -U_3 \\
 0 & 0 & 0 & 0& U_2 \\
 0 & 0 & 0 & 0& -U_1 \\
 1 & 0 & 0 & 0& 0 \\
 0 & 1 & 0 & 0& 0 \\
 0 & 0 & 1 & 0& 0 \\
 0 & 0 & 0 & 1& 0 
 \end{array} \right )_{\vert\tilde \Gamma_\nu} + {\cal O}_1(u)=
\sqrt{8\mu}\left (\begin{array}{ccccc}
 0 & 0 & 0 & 0& \nu_4 \\
 0 & 0 & 0 & 0& -\nu_3 \\
 0 & 0 & 0 & 0& \nu_2 \\
 0 & 0 & 0 & 0& -\nu_1 \\
 1 & 0 & 0 & 0& 0 \\
 0 & 1 & 0 & 0& 0 \\
 0 & 0 & 1 & 0& 0 \\
 0 & 0 & 0 & 1& 0 
 \end{array} \right )  + {\cal O}_1(u).
$$
Since $\norm{\nu}=1$, the rank of 
the matrix ${\cal J}(u,U)_{\vert\tilde \Gamma_\nu}$ is equal to 5 in a
neighbourhood of $(u,U)=(0,\sqrt{8\mu}\nu)$.
\hfill $\square$

\vskip 0.4 cm
It remains therefore to represent the motions on the
3-dimensional locally invariant manifolds $\tilde \Gamma_\nu$, and this 
will be done by defining from the function $W$ a suitable 
canonical transformation. Precisely, we consider the system:
\begin{eqnarray}
U_i &=& {\partial W\over \partial u_i}(u,\nu;E,\mu),\ \ i=1,\ldots ,4 \label{eqUbis}\\
n_i &=& {\partial W\over \partial \nu_i}(u,\nu;E,\mu),
\ \ i=1,\ldots ,4. \label{eqbetabis}  
\end{eqnarray}
which is well defined since the function $W$ can be differentiated
with respect to the variables $\nu_i$.
\vskip 0.4 cm
\noindent
    {\bf Inversion of the sub-system  (\ref{eqUbis}).} We first consider
    the sub-system formed by equations  (\ref{eqUbis}):
\begin{equation}
U_i = {\partial W\over \partial u_i}(u,\nu;E,\mu),\ \ i=1,2,3,4 .
\end{equation}
From (\ref{j4}), (\ref{Wfirst}) and the analyticity of $W$ with respect to
$u,\nu$, for any $\nu^*\in {\Bbb S}^3$  
we have the local inversion of the sub-system (\ref{eqUbis})
with respect to the variables $\nu$, in a neighbourhood
of $(u,\nu)=(0,\nu^*)$:
$$
\nu = \hat \nu (u,U;E,\mu)  ,
$$
and the functions $\hat \nu$ are analytic. As a matter of fact,
we have the stronger result:
\vskip 0,4 cm
\noindent
    {\bf Lemma 1.} {\it The sub-system  (\ref{eqUbis}) has a global
 analytic inversion:
$$
\nu = \hat \nu (u,U;E,\mu)  ,
$$
defined for $u,U$ so that $u$ is in some complex ball $B(d_0)$ and
for $U$ in the image of the map:
$$
\nu \longmapsto {\partial W\over \partial u}(u,\nu;E,\mu)
$$
with $\nu\in \Omega_{c_0}=\{ \nu: \norm{\norm{\nu}-1}< c_0\}$ with some
suitable $c_0,d_0$.
}
\vskip 0,4 cm
\noindent
    {\bf Proof of lemma 1.} We first proof that for fixed $E,
    \mu$, for all $u$ suitably
    close to $u=0$, and for suitably small $c_1$, the map:
    \begin{eqnarray}
    &\Psi_u: \Omega_{c_1} \longrightarrow {\Bbb R}^4&\cr
    &\nu \longmapsto {\partial W\over \partial u}(u,\nu;E,\mu)&
    \end{eqnarray}
    is injective. From (\ref{Wfirst}), we have the
    representation:
    $$
    \Psi_u(\nu) = \sqrt{8\mu}\ \nu +\psi_u(u,\nu;E,\mu)
    $$ 
    with $\psi_u(u,\nu;E,\mu)={\cal O}_2(u)$. 

    For arbitrary $k\geq 3$, we extend the map $\Psi_u$ to a map
    \begin{eqnarray}
    &\Psi^k_u: {\cal B}(1+c_1) \longrightarrow {\Bbb R}^4&\cr
      &\nu \longmapsto \Psi^k_u(\nu)& = \sqrt{8\mu}\nu +\phi^k(\norm{\nu})
      \psi_u(u,\nu;E,\mu)
    \end{eqnarray}
    where ${\cal B}(1+c_1)$ is the real ball centered at $\nu=0$
    of radius $1+c_1$ and
    $$
    \phi^k:  [0,1+c_1) \longrightarrow {\Bbb R}^4
    $$
    is a ${\cal C}^k$--smooth function such that
    $\phi^k(x)=1$ if $x\in [1-c_1/2,1+c_1)$, $\phi^k(x)=0$
      if $x\in [0,1-c_1]$, and in the interval $( 1-c_1,1-c_1/2)$
      increases smoothly and monotonically from $0$ to $1$. 
      For any fixed $k$, by restricting eventually the domain
      of $u$, we have that the map $\Psi^k_u$ is convex in the
      set ${\cal B}(1+c_1)$. Then, from a result on the global
      inversion of convex maps (see Theorem 4.2, page 137,
      of \cite{berger}), the map $\Psi^k_u$ is injective. But
      this implies that the also the map:
      \begin{eqnarray}
    &\Psi_u: \Omega_{c_1\over 2} \longrightarrow {\Bbb R}^4&\cr
    &\nu \longmapsto {\partial W\over \partial u}(u,\nu;E,\mu)&
    \end{eqnarray}
      is injective (in fact, if $\Psi_u(\nu')=\Psi_u(\nu'')$ with
      $\nu',\nu''\in \Omega_{c_1\over 2}$, then we have also
      $\Psi^k_u(\nu')=\Psi^k_u(\nu'')$ and therefore $\nu'=\nu''$)
      and therefore has the inverse:
      $$
        \Psi^{-1}_u: \Psi_u(\Omega_{c_1\over 2}) \longrightarrow
        \Omega_{c_1\over 2}  .
        $$
From the local inversion theorem the inverse map is analytic. \hfill $\square$
 \vskip 0.4 cm   
\noindent
{\bf The canonical transformation.} The inversion of the system  of equations (\ref{eqUbis}) provides the 
functions:
\begin{eqnarray}
\nu_i &=& \hat \nu_i (u,U;E,\mu)\cr
n_i &=& \hat n_i( u,U;E,\mu)\ \ i=1,2,3,4
\label{chi4}
\end{eqnarray}
which define a  canonical transformation:
$$
(n,\nu)=\chi_4(u,U)
$$
conjugating ${\cal K}_{\cal I}$ to the Hamiltonian:
$$
\hat {\cal K}(n,\nu)=\mu (\norm{\nu}^2-1)  .
$$
Therefore, the momenta $\nu_i$ are constants of motion and
the solutions $(u(s),U(s))$ of the Hamilton equations of
${\cal K}_{\cal I}$ are obtained from the inversion of:
$$
(n(0)+2\mu\ \nu(0)s,\nu(0))=\chi_4(u(s),U(s))  .
$$

\vskip 0.4 cm
\noindent
{\bf The bilinear relation.} From the identity:
$$
W(u,\nu;E,\mu) = W({\cal S}^0_\alpha u,{\cal S}^0_\alpha \nu;E,\mu) ,\hskip 1 cm 
\forall \alpha\in {\Bbb R}
$$
by differentiating both sides with respect to $\alpha$ and
computing in $\alpha=0$ we obtain:
$$
l\left ({\partial W\over \partial u}(u,\nu,E,\mu), u\right ) +
l\left ({\partial W\over \partial \nu}(u,\nu,E,\mu), \nu\right )=0
$$
and therefore we have $l(u,U)=0$ if and only if
$l(\hat n,\hat \nu)=0$. Consistently, $l(n,\nu)$ is 
a first integral of the Hamilton equations of $\hat {\cal K}(n,\nu)$.

\section{The first integrals in the space of the Cartesian variables}

In the previous section we have constructed four first integrals
$\hat \nu_i(u,U;E,\mu)$ 
of the KS Hamiltonian which are analytic in a neighbourhood  
of the collision set, represented in the space of coordinates
$u,U$ by:
$$
C=\{ (u,U)\in T^*{\Bbb R}^4: u=0,\ \ \Norm{U}=\sqrt{8\mu}\} .
$$
It is therefore interesting to know if, from the $\hat \nu_i$, it is
possible to construct first integrals $N_i(X,Y,Z,P_X,P_Y,P_Z)$ defined
in the Cartesian phase--space of the variables $(X,Y,Z,P_X,P_Y,P_Z)$
introduced in Section 2, eq. (\ref{planetoXYZ}). 

Following \cite{GL18}, we first show that from each $\hat \nu_i$
we construct a family of local first integrals defined
only in a neighbourhood of any point $(X,Y,Z,P_X,P_Y,P_Z)$,
with  $(X,Y,Z)$ in a neighbourhood of $(0,0,0)$; from this family,
we construct 2 first integrals (independent on the energy $E$)
which are globally defined in a complete neighbourhood of
$(X,Y,Z)=(0,0,0)$.  
\vskip 0.4 cm
\noindent
{\bf A phase-spaces projection.} We introduce a projection from the space:
$$
T^*{\Bbb R}^4_0=\{ (u,U) \in T^*{\Bbb R}^4: \norm{u}\ne 0\ \ ,\ \
l(u,U)=0\}
$$
to the Cartesian phase space of the variables $(X,Y,Z,P_X,P_Y,P_Z)$
introduced in Section 2, eq. (\ref{planetoXYZ}). We denote:
$$
(X,Y,Z,P_X,P_Y,P_Z)=\tilde \pi(u,U)
$$
where $(X,Y,Z)=\pi(u)$ and:
\begin{equation}
(P_X,P_Y,P_Z,0)={1\over 2\norm{u}^2}A(u)U  
  \label{projmomenta}
\end{equation}
\vskip 0.4 cm
\noindent
{\bf Local inversions of the phase-space projection.} We consider a
    local inversion of $(X,Y,Z)=\pi(u)$:
$$
\pi^{-1}: \ {\cal W} \longrightarrow  {\Bbb R}^4
$$
$$
(X,Y,Z) \longmapsto  u=\pi^{-1}(X,Y,Z)
$$
with ${\cal W}\subseteq {\mathbb R}^3\backslash 0$ open set, and define:
$$
(u,U)=\chi(X,Y,Z,P_X,P_Y,P_Z)
$$
where $u=\pi^{-1}(X,Y,Z)$ and, from  (\ref{projmomenta}):
$$
U=2 A(u)^T (P_X,P_Y,P_Z,0)  .
$$
We introduce the matrix:
\begin{equation}\label{omega}
\Omega = 
\left ( \begin{array}{cccc}
0 & 0 & 0 & -1 \\
0 & 0 &  1 & 0 \\
0 & -1 & 0 & 0 \\
1 & 0 & 0  & 0
 \end{array} \right ) ,
\end{equation} 
so that $l(u,U)=u\cdot \Omega U$. We notice that we have:
$$
l(u,U)=2 u\cdot \Omega A(u)^T (P_X,P_Y,P_Z,0)=
2 (A(u)\Omega^T u)\cdot (P_X,P_Y,P_Z,0)=0  ,
$$
since $A(u)\Omega^T u$ is a four dimensional vector with
only the fourth component different from zero. Therefore, for any
choice of $\pi^{-1}$, the phase-space local inversion
$\chi$ is well defined in $T^*{\mathbb R}^4_0$. 
\vskip 0.4 cm
\noindent
    {\bf An atlas of local inversions.} Following \cite{GL18} (where
    a similar result is proved between the Cartesian state-space
    with coordinates $x,y,z,\dot x,\dot y,\dot z$ and the state-space
    of the KS variables $u,u'$) we define an atlas of two local inversions of the map $\pi$ 
defined in ${\Bbb R}^3\backslash (0,0,0)$: 
\vskip 0.2 cm
\noindent
{\bf Lemma 2.} {\it Consider the maps
$$
\pi^{-1}_{-}: D_-={\Bbb R}^3\backslash\{ (X,0,0): X\geq 0\} \longrightarrow {\Bbb R}^4
$$
$$
\pi^{-1}_{+}: D_+={\Bbb R}^3\backslash\{ (X,0,0): X\leq 0\} 
\longrightarrow {\Bbb R}^4
$$
defined by
$$
\pi^{-1}_{-}(X,Y,Z)=  \left ( {Y\over \sqrt{2(r-X)}},
{\sqrt{r-X}\over \sqrt{2}},0, {Z \over \sqrt{2(r-X)}}\right )
$$
$$
\pi^{-1}_{+}(X,Y,Z)=\left (  {\sqrt{r+X}\over \sqrt{2}},
{Y\over \sqrt{2(r+X)}},{Z \over \sqrt{2(r+X)}},0\right )  ,
$$
where $r=\sqrt{X^2+Y^2+Z^2}$, as well as their phase-space 
extensions:
$$
\chi_{\pm }: ({\Bbb R}^3\backslash 0)\times {\Bbb R}^3 
\longrightarrow T^*{\mathbb R}^4_0 
$$
$$
(X,Y,Z,P_X,P_Y,P_Z) \longmapsto (u,U)=\chi_{\pm }(X,Y,Z,P_X,P_Y,P_Z)
$$
defined by:
$$
\chi^{-1}_{-}(X,Y,Z,P_X,P_Y,P_Z)=\left ( \pi^{-1}_{-}(X,Y,Z) , 2 
A(\pi^{-1}_{-}(X,Y,Z) )^T (P_X,P_Y,P_Z,0)\right )
$$
$$
\chi^{-1}_{+}(X,Y,Z,P_X,P_Y,P_Z)=\left ( \pi^{-1}_{+}(X,Y,Z) , 2 
A(\pi^{-1}_{+}(X,Y,Z) )^T (P_X,P_Y,P_Z,0)\right )  .
$$
Then, for every $(X,Y,Z,P_X,P_Y,P_Z)$ in the domain of $\chi^{-1}_{-}$ we have:
$$
(X,Y,Z,P_X,P_Y,P_Z)= \chi \circ \chi^{-1}_{-}(X,Y,Z,P_X,P_Y,P_Z) ,
$$
for every $(X,Y,Z,P_X,P_Y,P_Z)$ in the domain of $\chi^{-1}_{+}$ we have:
$$
(X,Y,Z,P_X,P_Y,P_Z)= \chi\circ \chi^{-1}_{+}(X,Y,Z,P_X,P_Y,P_Z) ,
$$
and, for every $(X,Y,Z,P_X,P_Y,P_Z)$ in the intersection of
the domains of $\chi^{-1}_{-}$ and $\chi^{-1}_{+}$ 
exists $\alpha\in {\Bbb R}$ (depending only on $(X,Y,Z)$) such that,
by denoting 
$$
(u_{\pm},U_{\pm})= \chi^{-1}_{\pm}(X,Y,Z,P_X,P_Y,P_Z)  ,
$$
we have:
\begin{equation}
u_+= {\cal S}^0_\alpha u_-\ \ ,\ \ U_+= {\cal S}^0_\alpha U_- \ \ .
\label{transition}
\end{equation}
}
\vskip 0.4 cm
\noindent
{\bf Proof of Lemma 2.} We prove that indeed we have 
$U_+= {\cal S}^0_\alpha U_-$. Since $u_+={\cal S}^0_\alpha u_-$, we have:
$$
U_+=2 A(u_+)^T(P_X,P_Y,P_Z)=2 A({\cal S}^0_\alpha u_- )^T(P_X,P_Y,P_Z,0)
$$
$$
={\cal S}^0_\alpha A(u_-)^T(P_X,P_Y,P_Z,0)={\cal S}^0_\alpha U_- .
$$
\hfill $\square$

\vskip 0.4 cm
\noindent
{\bf Cartesian representatives of the $\hat \nu_i,\hat n_i$.} 
Let us fix $E,\mu$, and consider the set
${\cal D}_E\subseteq (D\backslash 0)\times {\Bbb R}^3$ where 
$D$ is a suitable small neighbourhood of $(0,0,0)$ and
for any $ (X,Y,Z,P_X,P_Y,P_Z)\in {\cal D}_E$ we have  
$H(X,Y,Z,P_X,P_Y,P_Z)=E$. In the sets:
$$
{\cal D}_E^{\pm}= {\cal D}_E \cap (D_{\pm}\times {\Bbb R}^3)
$$
we define:
$$
\tilde \nu_{i,\pm}(X,Y,Z,P_X,P_Y,P_Z)= 
\hat \nu_i(\chi^{-1}_\pm (X,Y,Z,P_X,P_Y,P_Z);E,\mu)  
$$
$$
\tilde n_{i,\pm}(X,Y,Z,P_X,P_Y,P_Z)= 
\hat n_i(\chi^{-1}_\pm (X,Y,Z,P_X,P_Y,P_Z);E,\mu)  .
$$
Since these functions are constructed using the local inversions
$\chi^{-1}_\pm$, they satisfy the identity:
$$
l(\tilde n_{\pm}(X,Y,Z,P_X,P_Y,P_Z),\tilde \nu_{\pm}(X,Y,Z,P_X,P_Y,P_Z))=0 ,
$$
and since they are constructed from the solutions of the Hamilton-Jacobi
equation on the zero energetic level of the KS Hamiltonian, they 
also satisfy:
$$
\norm{\tilde \nu_{i,\pm}(X,Y,Z,P_X,P_Y,P_Z)}=1 .
$$
Let us denote by:
$$
(u_{\pm},U_{\pm})= \chi^{-1}_{\pm}(X,Y,Z,P_X,P_Y,P_Z) 
$$
the pre-images, by $\alpha$ the angle such that:
$$
u_+= {\cal S}^0_\alpha u_-\ \ ,\ \ U_+= {\cal S}^0_\alpha U_- ,
$$
and:
$$
\nu_+=\hat \nu(u_+,U_+;E,\mu)\ \ , \nu_-=\hat \nu(u_-,U_-;E,\mu)  
$$
$$
n_+= \hat n(u_+,U_+;E,\mu)\ \ , n_-=\hat n(u_-,U_-;E,\mu)  .
$$
We prove:
\begin{equation}
\nu_+= {\cal S}^0_\alpha \nu_- \ \ ,\  \ n_+= {\cal S}^0_\alpha n_-
\label{nupiumeno}
\end{equation}
Since:
$$
U_+ = {\partial W\over \partial u}(u_+,\nu_+;E,\mu)
$$
$$
U_-={\partial W\over \partial u}(u_-,\nu_-;E,\mu)
$$
we have:
\begin{equation}
{\partial W\over \partial u}(u_-,\nu_-;E,\mu)=({\cal S}^0_\alpha)^T 
{\partial W\over \partial u}({\cal S}^0_\alpha u_-,\nu_+;E,\mu)  .
\label{walpha}
\end{equation}
We use the previous equation to establish the relation between $\nu_+$ and
$\nu_-$.
\vskip 1 cm
Let us consider the complete integral $W(u,\nu;E,\mu)$ of the Hamilton-Jacobi
equation:
$$
{\cal K}_{\cal I}\left (u,{\partial W\over \partial u}(u,\nu;E,\mu)\right )
=\mu (\norm{\nu}^2-1)
$$
defined in Section 5. In particular, for any $\nu$ in a suitable
small neighbourhood of the sphere $\norm{\nu}=1$, the function 
$W(u,\nu;E,\mu)$ is analytic in a  neighbourhood of $u=0$ and,
if also $u\cdot \nu=0$, we have: 
$$
W(u,\nu;E,\mu)=0 \ \ .
$$
For any $\alpha\in {\Bbb R}$, let us define the function:
$$
\hat W_\alpha(u,\nu;E,\mu) = W({\cal S}^0_\alpha u,{\cal S}^0_\alpha \nu;E,\mu)  .
$$
We prove:
$$
W(u,\nu;E,\mu) = \hat W_\alpha(u,\nu;E,\mu)  .
$$
In fact, since ${\cal S}^0_\alpha$ acts as a symmetry for the
Hamiltonian ${\cal K}_{\cal I}$,
$$
{\cal K}_{\cal I}({\cal S}^0_\alpha u,{\cal S}^0_\alpha U)=
{\cal K}_{\cal I}(u,U)  ,
$$
we have:
$$
{\cal K}_{\cal I}\left (u,{\partial \hat W_\alpha
  \over \partial u}(u,\nu;E,\mu)\right )=
{\cal K}_{\cal I}\left (u, ({\cal S}^0_\alpha)^T
{\partial W \over \partial u}({\cal S}^0_\alpha u,{\cal S}^0_\alpha\nu;E,\mu)\right )
$$
$$
={\cal K}_{\cal I}\left ({\cal S}^0_\alpha u, {\partial W \over \partial u}({\cal S}^0_\alpha u,{\cal S}^0_\alpha\nu;E,\mu)\right )=
\mu (\norm{{\cal S}^0_\alpha\nu}^2-1)=\mu (\norm{\nu}^2-1)
$$
and therefore $\hat W_\alpha(u,\nu;E,\mu)$ is a solution of the
Hamilton-Jacobi equation. Also,  $\hat W_\alpha(u,\nu;E,\mu)=0$ on the
hyperplane $u\cdot \nu=0$. Therefore $W,\hat W_\alpha$ are both  
solutions of the same Hamilton-Jacobi equation; they are both analytic in a 
common neighbourhood of $u=0$; they both vanish on the same
hyperplane. Therefore, they coincide in their common domain: 
$$
W(u,\nu;E,\mu) = W({\cal S}^0_\alpha u,{\cal S}^0_\alpha \nu;E,\mu)  ,
$$
and in particular we have the identity:
$$
{\partial W\over \partial u}(u,\nu;E,\mu)=
({\cal S}^0_\alpha)^T {\partial W\over \partial u}({\cal S}^0_\alpha u,
{\cal S}^0_\alpha \nu;E,\mu)  .
$$
Therefore, from eq. (\ref{walpha}), we have:
$$
{\partial W\over \partial u}({\cal S}^0_\alpha u_-,
{\cal S}^0_\alpha \nu_-;E,\mu) =
{\partial W\over \partial u}({\cal S}^0_\alpha u_-,\nu_+;E,\mu)
$$
and from Lemma 1: $\nu_+={{\cal S}^0_\alpha}\nu_{-}$. Finally, we have:
$$
n_-= {\partial W\over \partial \nu}({u_-},{\nu_-};E,\mu) =
({{\cal S}^0_\alpha})^T  {\partial W\over \partial \nu}({{\cal S}^0_\alpha}
{u_-},{{\cal S}^0_\alpha}{\nu_-};E,\mu)
$$
$$
=({{\cal S}^0_\alpha})^T {\partial W\over \partial \nu}({u_+},{\nu_+};E,\mu)=
({{\cal S}^0_\alpha})^Tn_+  .
$$
\vskip 0.4 cm
\noindent
    {\bf From local to global first integrals.} The functions
    $\tilde \nu_{\pm}$, $\tilde n_{\pm}$ constructed above
    indeed depend on the chart  ${\cal D}_E^\pm$, and therefore are not
    globally defined in ${\cal D}_E$.  We here aim to construct, from
    the functions     $\nu(u,U,E),n(u,U,E)$, first
    integrals in the Cartesian coordinates  which are globally
    defined in ${\cal D}_E$. First of all, we consider the dynamics
    in the $\nu,n$ variables:
    $$
    \nu_i(s)=\nu_i(0)\ \ ,\ \ n_i(s)=n_i(0)+2\mu \nu_i(0)s
    $$
    and we notice that the functions:
    $$
    N_X=\nu_1 n_4-\nu_4 n_1
    $$
    $$
    N_Y={1\over 2}(\nu_1 n_3-n_1 \nu_3+n_2\nu_4-n_4\nu_2)
    $$
    $$
    N_Z={1\over 2}(\nu_1 n_2-n_1 \nu_2+n_4\nu_3-n_3\nu_4)
    $$
    are first integrals. Since they are all invariant
    by composition with the map $(n,\nu)\mapsto ({\cal S}^0_\alpha n,
    {\cal S}^0_\alpha \nu)$ for any $\alpha$, their local representatives:
    $$
    N_X^{\pm}(X,Y,Z,P_X,P_Y,P_Z)=({\tilde \nu}^\pm_1 {\tilde n}^\pm_4
        -{\tilde \nu}^\pm_4 {        \tilde n}^\pm_1)(X,Y,Z,P_X,P_Y,P_Z)
    $$
    $$
    N_Y^{\pm}(X,Y,Z,P_X,P_Y,P_Z)={1\over 2}({\tilde \nu}^\pm_1 {\tilde n}^\pm_3
        -{\tilde \nu}^\pm_3 { \tilde n}^\pm_1 +{\tilde \nu}^\pm_4 {\tilde n}^\pm_2
        -{\tilde \nu}^\pm_2 { \tilde n}^\pm_4)(X,Y,Z,P_X,P_Y,P_Z)   
    $$
    $$
    N_Z^{\pm}(X,Y,Z,P_X,P_Y,P_Z)={1\over 2}({\tilde \nu}^\pm_1 {\tilde n}^\pm_2
        -{\tilde \nu}^\pm_2 { \tilde n}^\pm_1 +{\tilde \nu}^\pm_4 {\tilde n}^\pm_3
        -{\tilde \nu}^\pm_3 { \tilde n}^\pm_4)(X,Y,Z,P_X,P_Y,P_Z)   
    $$
    satisfy, for all $(X,Y,Z,P_X,P_Y,P_Z)\in {\cal D}_E^+\cap{\cal D}_E^- $:
    $$
    N_X^{+}(X,Y,Z,P_X,P_Y,P_Z)=N_X^{-}(X,Y,Z,P_X,P_Y,P_Z)\ \ ,\ \ 
    $$
    $$
    N_Y^{+}(X,Y,Z,P_X,P_Y,P_Z)=N_Y^{-}(X,Y,Z,P_X,P_Y,P_Z)\ \ ,\ \ 
    $$
    $$
    N_Z^{+}(X,Y,Z,P_X,P_Y,P_Z)=N_Z^{-}(X,Y,Z,P_X,P_Y,P_Z)\ \ ,\ \ ,
    $$
    and therefore are the local representatives of a functions
${\cal N}_X,{\cal N}_Y,{\cal N}_Z$ globally defined in ${\cal D}_E$. Now
we allow $E$ change in a small neighbourhood ${\cal E}$ of a given $E_*$, and 
we consider the set of three first integrals:
$$
\Big (\ H \ \ , \ \ {\cal N}^2:={\cal N}_X^2+{\cal N}_Y^2+{\cal N}_Z^2\ \ ,\ \ 
{\cal N}_Z\ \Big )
$$
defined in $\cup_{E\in {\cal E}}{\cal D}_E$. We have the following:
\vskip 0.4 cm
\noindent
{\bf Theorem.} {\it The set of first integrals $(H,{\cal N}^2,{\cal N}_Z)$ 
is complete.} 
\vskip 0.4 cm
\noindent
{\bf Proof.} Let us prove that $(H,{\cal N}^2,{\cal N}_Z)$ are independent
in a set $\cup_{E\in {\cal E}}{\cal D}_E$. 
We first prove that
${\cal N}^2,{\cal N}_Z$ are independent on $E$, by showing that they are
not constant on the energy levels $H(X,Y,Z,P_X,P_Y,P_Z)=E$.

For any arbitrary small $\epsilon$, in the set
\begin{equation}
\{(X,Y,Z,P_X,P_Y,P_Z)\in\cup_{E\in {\cal E}}{\cal D}_E, \ \
0<\Norm{(X,Y,Z)}<\epsilon^2\}
\label{setepsilon}
\end{equation}
we have:
\begin{eqnarray}
{\cal N}_X &=& P_Y Z-P_Z Y+D_X\cr
{\cal N}_Y &=& P_X Z-P_Z X+D_Y \cr
{\cal N}_Z &=& P_X Y-P_Y X+D_Z\cr
{\cal N}^2 &=& (P_Y Z-P_Z Y)^2+ (P_X Z-P_Z X)^2+( P_X Y-P_Y X)^2+D^2
\end{eqnarray}
where the functions $D_X,D_Y,D_Z$ have sup-norm bounded
by order $\epsilon^3$ and $D^2$ bounded by order $\epsilon^4$ in
the set (\ref{setepsilon}). Therefore, if we fix the value of $E$
and one between ${\cal N}_Z,{\cal N}^2$, the third integral is
not constant in the level set of the first two. 

Let us now compute the Poisson brackets. Since ${\cal N}_Z,{\cal N}^2$
are first integrals for the Hamilton equations of $H$, we have:
$$
\{H,{\cal N}_Z\}=0\ \ ,\ \ \{H,{\cal N}^2\}=0 \ \ .
$$
It remains to compute the Poisson bracket $\{{\cal N}^2,{\cal N}_Z\}$. 
Let us denote by $\hat q(u),\hat p(u,U)$ the functions defined
by:
$$
\hat q(u) =\pi(u)\ \ ,\ \ (\hat p(u,U),0)={1\over 2 \norm{u}^2}A(u)U  .
$$
We notice the remarkable property of the Poisson brackets:
\begin{equation}\label{ppqp}
\{ \hat q_i, \hat  p_j\} = \delta_{ij} ,\ \ \{\hat  q_i,\hat  q_j\}=0 ,\ \ \{\hat p_i,\hat p_j\}=
l(u,U)\phi_{ij}(u,U)\ \ ,i,j=1,2,3 .
\end{equation}
and from:
\begin{equation}
\{ N^2,N_Z\} = l(n,\nu)a(n,\nu)  ,
\label{atipicpp}
\end{equation}
we prove $\{{\cal N}^2,{\cal N}_Z\}=0$. In fact,
since ${\cal N}^2,{\cal N}_Z$ they are invariant
    by composition with the map $(n,\nu)\mapsto ({\cal S}^0_\alpha n,
    {\cal S}^0_\alpha \nu)$ for any $\alpha$, we have:
    $$
    \hat N_Z(u,U;E)=N_z(\chi_4(u,U;E))={\cal N}_Z(\hat q(u,U),\hat p(u,U))
    $$
    $$
    \hat N^2(u,U;E)=N^2(\chi_4(u,U;E))={\cal N}^2(\hat q(u,U),\hat p(u,U)) .
    $$
    By denoting with ${\mathbb E}_k$ the standard symplectic matrix
    of ${\mathbb R}^{2k}$ and $q=(X,Y,Z),p=(P_X,P_Y,P_Z)$,  
    we have:
    $$
    \{\hat N^2(u,U;E),\hat N_Z(u,U;E)\} =
    \left ( {\partial N^2\over \partial u},{\partial N^2\over \partial U}\right )\cdot \left ({\mathbb E}_4 \left ( {\partial N_Z\over \partial u},{\partial N_Z\over \partial U}\right )\right )
    $$
    $$
    =\left ( {\partial {\cal N}^2\over \partial q},
          {\partial {\cal N}^2\over \partial p}\right )\cdot \left (J(u,U)^T {\mathbb E}_4J(u,U) \left ( {\partial {\cal N}_Z\over \partial q},{\partial {\cal N}_Z\over \partial p}\right )\right )
$$
          where $J$ is the $8\times 6$ Jacobian matrix of $(\hat q(u),\hat p(u,U))$. From the Poisson brackets (\ref{ppqp}) we notice that the
          $6\times 6$ matrix $J(u,U)^T {\mathbb E}_4J(u,U)$
          is not identically equal to ${\mathbb E}_3$, but when it
          is computed on $(u,U)$ satisfying $l(u,U)=0$ we have:
          $J(u,U)^T {\mathbb E}_4J(u,U)={\mathbb E}_3$. But, from
          (\ref{atipicpp}), for $l(u,U)=0$ we also have
          $\{\hat N^2(u,U;E),\hat N_Z(u,U;E)\}=0$. Finally,
          since have identified the preimages of $\hat q,\hat p$ satisfying
          $l(u,U)=0$, we have:  $\{{\cal N}^2,{\cal N}_Z\}=0$. \hfill $\square$

\section{Appendix 1: a revisitation of the integrability of the LC Hamiltonian in a 
neighbourhood of the collision singularities}

Let us consider the Hamiltonian of the planar circular 
restricted three-body problem in the planetocentric reference
frame (see (\ref{hamplanetocentric}) for comparison):
$$
H_2(X,Y,P_X,P_Y)= {P_X^2+P_Y^2\over 2} + P_X Y -P_Y X-{\mu\over
\sqrt{X^2+Y^2}}
$$
\begin{equation}
-(1-\mu)\left ({1 \over \sqrt{(X+1)^2+Y^2}}-1+X\right )
-(1-\mu)-{(1-\mu)^2\over 2}  .
\label{hamplanetocentric2}
\end{equation}
Following Levi-Civita we first define the canonical transformation: 
$$
(X,Y,P_X,P_Y)={\cal Y}(u_1,u_2,U_1,U_2)
$$
where:
$$
X=u_1^2-u_2^2 \ \ ,\ \ Y=2u_1u_2
$$
represents the equations (\ref{regcla}), (\ref{regclbb}) in the planetocentric
reference frame, and:
$$
P_X= {U_1u_1-U_2u_2\over 2 \norm{u}^2} \ \ , P_Y={U_1u_2+U_2 u_1\over 
2 \norm{u}^2}
$$
the canonical extension to the momenta $U_1,U_2$. The transformation 
${\cal Y}$ conjugates $H_2$ to the Hamiltonian:
$$
K_2(u_1,u_2,U_1,U_2)= {1\over 8\norm{u}^2}\left (U_1+2\norm{u}^2 u_2\right )^2
+ {1\over 8\norm{u}^2}\left (U_2-2\norm{u}^2 u_1\right )^2
-{1\over 2}\norm{u}^4-{\mu\over \norm{u}^2}
$$
\begin{equation}
-(1-\mu)\left ({1 \over \sqrt{1+2(u_1^2-u_2^2)+\norm{u}^4}}-1+u_1^2-u_2^2\right )
-(1-\mu)-{(1-\mu)^2\over 2}  .
\label{kamplanetocentric2}
\end{equation}
To remove the singularity at $u=0$ we perform the iso-energetic reduction:
for any value $E$ of the Hamiltonian, we introduce the LC Hamiltonian:
$$
{\cal K}_2(u,U;E)=\norm{u}^2(K_2(u,U)-E)={1\over 8}\left (U_1+2\norm{u}^2 u_2\right )^2+ {1\over 8}\left (U_2-2\norm{u}^2 u_1\right )^2
$$
$$
-{1\over 2}\norm{u}^6-\mu-\norm{u}^2 \left (E+(1-\mu)+{(1-\mu)^2\over 2}\right )
$$
\begin{equation}
-(1-\mu)\norm{u^2}
\left ({1 \over \sqrt{1+2(u_1^2-u_2^2)+\norm{u}^4}}-1+u_1^2-u_2^2\right )  .
\label{LC-hamiltonian}
\end{equation}
The LC Hamiltonian is regular at $u=(0,0)$, and the solutions 
$(u(s),U(s))$ 
of the Hamilton equations of ${\cal K}_2(u,U)$:
$$
{d\over ds} u_i = {\partial {\cal K}_2   \over \partial U_i} \ \ ,\ \ 
{d\over ds} U_i = - {\partial {\cal K}_2   \over \partial u_i} ,
$$
with initial conditions satisfying $u(0)\ne 0$ and ${\cal K}_2(u,U)=0$, 
are conjugate, in a neighbourhood of $s=0$, to solutions $(X(t),Y(t),P_x(t),P_y(t))$ of the Hamilton equations of (\ref{hamplanetocentric2}) 
after the replacement of the proper time $s$ with the time $t$ through the
formula:
\begin{equation}\label{propertime}
t(s)=\int_0^s \norm{u(\tau)}^2d\tau .
\end{equation}
A complete integral of the Hamilton-Jacobi equation:
\begin{equation}
{\cal K}_2\left (u ,{\partial W\over \partial u};E\right )= \kappa ,
\label{HJ-K2}
\end{equation}
is a family of solutions of (\ref{HJ-K2}) depending on two parameters, 
satisfying the usual non-transversality property. We identify 
the two parameters in\footnote{In \cite{LC1906} only the case $\kappa =0$ was considered, which actually 
is the only value which grants the conjugation between the solutions
of the regularized and non-regularized equations.} $\kappa$ and, following
T. Levi-Civita, an angle
$\alpha$ related to the rotations of the plane $u_1,u_2$. We have the following proposition: there
exists a family of solutions $W(u_1,u_2,\alpha,\kappa )$ of 
the Hamilton-Jacobi equation (\ref{HJ-K2}), defined for 
$\alpha \in {\Bbb S}^1$, $\kappa$ in a neighbourhood of $\kappa=0$, 
and analytic for $u_1,u_2$ in neighbourhhod of $(0,0)$ of radius
$\sigma>0$, with $\sigma$ depending from $E$ and $\mu$. 
Moreover:
\begin{itemize}
\item[(i)] the Taylor series of $W$:
$$
W= \sum_{n_1,n_2}W_{n_1,n_2}(\alpha,\mu,\kappa,E)u_1^{n_1}u_2^{n_2}
$$
has coefficients periodic in $\alpha$, which can be computed iteratively
to any order $n_1+n_2$. In particular, we have:
\begin{equation}
W= \sqrt{8 (\mu+\kappa)}(u_1\cos \alpha +u_2\sin\alpha )+{\cal O}_2(u_1,u_2) .
\label{w2}
\end{equation}

\item[(ii)] By denoting:
$$
j_2(u_1,u_2,\alpha,\kappa)= \det \left ( 
\begin{array}{cc}
{\partial W\over \partial u_1\partial \alpha} & 
{\partial W\over \partial u_1\partial \kappa} \\
{\partial W\over \partial u_2\partial \alpha} & 
{\partial W\over \partial u_2\partial \kappa}
 \end{array} \right ) ,
$$
from (\ref{w2}), we obtain $j_2(0,0,\alpha,\kappa)=4$. Therefore
we have $j_2(0,0,\alpha,\kappa)\ne 0$ in a neighbourhood of $(u_1,u_2)=(0,0)$,
uniformly in $\alpha$ and for $\kappa$ in a small neighbourhood of $\kappa=0$. 
\end{itemize}

As a consequence, the system:
\begin{eqnarray}
U_1&=& {\partial W\over \partial u_1}(u_1,u_2,\alpha,\kappa) \cr
U_2&=& {\partial W\over \partial u_2}(u_1,u_2,\alpha,\kappa) \cr
\beta &=& - {\partial W\over \partial \alpha}(u_1,u_2,\alpha,\kappa) \cr
K &=& s + {\partial W\over \partial \kappa}(u_1,u_2,\alpha,\kappa)
\label{LC-system}
\end{eqnarray}  
defines by inversion a $s$-dependent canonical transformation
$$
(\alpha , \kappa , \beta , K)= \chi_2 (s, \alpha , \kappa , \beta , K),
$$
conjugating the Hamiltonian ${\cal K}_2$ to the zero-value Hamiltonian
${\hat  {\cal K}}_2 (s, \alpha , \kappa , \beta , K)=0$. 
In particular, by selecting the value $\kappa=0$, equations (\ref{LC-system})
provide the solution to the problem of planar close encounters.   
 \vskip 0.4 cm
 The proof of the existence of the complete integral $W$ has been done in
 \cite{LC1906} as follows. Consider the canonical transformation
$$
u = {\cal R}_\alpha \tilde u \ \ ,\ \ U= {\cal R}_\alpha \tilde U
$$
where
\begin{equation}
{\cal R}_\alpha = 
\left ( \begin{array}{cc}
\cos\alpha & -\sin\alpha \\
\sin\alpha & \cos\alpha
 \end{array} \right ) ,
\end{equation}
conjugating ${\cal K}_2(u,U;E)$ to the Hamiltonian:
$$
{\tilde {\cal K}}_2(\tilde u,\tilde U;\alpha,E)=
{1\over 8}\left (\tilde U_1+2\norm{\tilde u}^2 \tilde u_2\right )^2+ {1\over 8}\left (\tilde U_2-2\norm{\tilde u}^2 \tilde u_1\right )^2
$$
$$
-{1\over 2}\norm{\tilde u}^6-\mu-\norm{u}^2 \left (E+(1-\mu)+{(1-\mu)^2\over 2}\right )
$$
\begin{equation}
-(1-\mu)\norm{\tilde u}^2
\left ({1 \over \sqrt{1+2(\tilde u_1^2-\tilde u_2^2)\cos 2\alpha+\norm{\tilde u}^4}}-1+(\tilde u_1^2-\tilde u_2^2)\cos 2\alpha\right )  ,
\label{LC-tildehamiltonian}
\end{equation}
and look for a particular solution $\tilde W(\tilde u,\alpha,\kappa)$ of 
the Hamilton--Jacobi equation:
$$
{\tilde K}_2\left (\tilde u,{\partial \tilde W\over \partial u};\alpha,E
\right )=\kappa
$$
satisfying:
\begin{equation}
\tilde W(0,\tilde u_2,\alpha,\kappa)=0
\label{ckcondition2}
\end{equation}
for all $u_2$ in a neighbourhood of $0$. The existence of a solution
to this problem which is analytic in a neighbourhood of $\tilde u=0$, 
(with a common analyticity radius to all $\alpha$)   
is quoted in \cite{LC1906} as a consequence on a general result about 
the regularity of the solutions of first order PDE, which we identify in 
the Cauchy--Kowaleski theorem (see \cite{CH}, and the Appendix). 
The complete integral $W(u;\alpha ,\kappa)$ is
then defined by:
$$
W= \tilde W({\cal R}^T_\alpha u;\alpha,\kappa)  .
$$
As it is usual in the Cauchy--Kowaleski theorem, the coefficients of the
series expansion of $W$ in $u_1,u_2$
can be computed iteratively up to any arbitrary order.

\section{Appendix 2: the Cauchy-Kowaleski theorem}\label{appendixck}

We consider the first order PDE:
\begin{equation}
{\partial W\over \partial q_1}=F\left (q_1,q_2,\ldots ,q_n,
{\partial W\over \partial q_2},\ldots ,{\partial W\over \partial q_n}\right )
\label{ckgenerico}
\end{equation}
where $F(q_1,\ldots ,q_n,p_1,\ldots ,p_{n-1})$ is analytic 
in a neighbourhood of $q=(q_1,\ldots ,q_n)=0$, $(p_1,\ldots ,p_{n-1})=0$. 
We call the plane $q_1=0$ the initial plane in the space of the 
variables $q$; then, we consider the Cauchy's problem of finding a solution 
$W(q)$ of the PDE (\ref{ckgenerico}) satisfying the given initial condition:
\begin{equation}
W(0,q_2,\ldots ,q_n)=\phi(q_2,\ldots ,q_n)
\label{condin}
\end{equation}
in a suitable neighbourhood of $(q_2,\ldots ,q_n)=(0,\ldots ,0)$, where 
$\phi$ is a given function analytic in a neighbourhood of $(q_2,\ldots ,q_n)=(0,\ldots ,0)$. The Cauchy-Kowaleski 
theorem states (see for example \cite{CH}) that the Cauchy problem has a 
unique solution analytic in a suitable small neighbourhood of 
$q=0$. We will continue our discussion in the case which is useful for our purposes,
defined by the special choice of the initial condition:
$$
\phi(q_2,\ldots ,q_n)=0 .
$$
The proof is obtained by constructing first a formal series expansion:
\begin{equation}
W= \sum_{i_1,\ldots ,i_n\geq 0}c_{i_1,\ldots ,i_n}q_1^{i_1}\cdots q_n^{i_n}
\label{seriesck}
\end{equation}
for the solution as follows. 
From: $W(0,q_2,\ldots ,q_n)=\phi(q_2,\ldots ,q_n)=0$ we immediately 
obtain:
$$
{\partial^{i_2} \over \partial q^{i_2}_{j_2}}\cdots 
{\partial^{i_n} \over \partial q^{i_n}_{j_n}}W(0,q_2,\ldots ,q_n)=0\ \ , 
j_2,\dots ,j_n=2,\ldots ,n
$$
for all $i_2,\ldots ,i_n \geq 1$; correspondingly,
we have $c_{0,i_2,\ldots ,i_n}=0$. The coefficients
with $i_1\ne 0$ are computed iteratively  on the order
$i_1+\ldots +i_n$ by differentiating (\ref{ckgenerico}) 
 and by computing the result at $q=0$. 

Then, the proof of the absolute convergence of the 
expansion (\ref{seriesck}) in a neighbourhood of $q=0$ is obtained by
using the method of majorants. To apply the method, one first observes 
that the terms $c_{i_1,\ldots ,i_n}$ computed as indicated above  
can be represented as polynomials of the terms of the Taylor
expansions of $F$ and $\phi$ at $q=0$, and the coefficients of 
these polynomials are non negative numbers. By exploiting this 
property one constructs a PDE problem whose solution 
can be given explicitly (and so its analyticity can be directly checked) 
and is a majorant of $W$. For the purposes of our paper, it is crucial
to remark that all the differential equations whose solution have the
same majorant converge in a common domain of $q=0$.   
\vskip 0,4 cm
\noindent
    {\bf Acknowledgements.} This research has been supported by 
ERC project 677793 Stable and Chaotic Motions in the Planetary Problem.
The author F.C. acknowledges also the project MIUR-PRIN  2017S35EHN titled 
``Regular and stochastic behaviour in dynamical systems". The author M.G. 
acknowledges also the project MIUR-PRIN 20178CJA2B titled ``New
frontiers of Celestial Mechanics: theory and applications".

\end{document}